\documentclass[11pt,letterpaper]{article}

\usepackage{jheppub}
\usepackage{braket}
\usepackage{amsmath}
\usepackage{amsfonts}
\usepackage{amssymb}
\usepackage{graphicx}
\usepackage{epsfig}
\usepackage{color}
\usepackage{hyperref}
\usepackage{enumitem}
\usepackage{multirow}
\usepackage{todonotes}
\usepackage{cleveref}
\usepackage{dsfont}
\usepackage{bm} 
\usepackage{cancel}
\usepackage{tikz}
\usepackage{dsfont}
\usepackage{braket}
\usepackage{float}
\usepackage{slashed}
\usepackage[caption=false]{subfig}
\usepackage{upgreek}
\usepackage{tabularray}
\usetikzlibrary{decorations.pathreplacing}
\usepackage[normalem]{ulem}
\usepackage[export]{adjustbox}
\usepackage{cleveref}
\usepackage{relsize}
\usepackage{scalerel}[2016/12/29]

\def\nn{\nonumber}
\def\d{\text{d}}
\def\modeone{k_{A}}
\def\modetwo{k_{B}}
\def\vmodeone{\vec{k}_{A}}
\def\vmodetwo{\vec{k}_{B}}
\def\vx{\vec{x}}
\def\vy{\vec{y}}

\def\vk{\vec{k}}

\def\vq{\vec{q}}
\def\hphi{\hat{\phi}}

\def\hpi{\hat{\pi}}
\def\hrho{\hat{\rho}}
\newcommand{\hc}{\hat{c}^{\phantom{\dagger}}}
\newcommand{\hcdagg}{\hat{c}^{\dagger}}
\def\del{\partial}

\renewcommand{\thetable}{\arabic{table}}

\newcommand{\inn}{\ensuremath{\mathbin{\raisebox{0.1ex}{\scalebox{0.6}{\ensuremath{\in}}}}}}

\newcommand{\Lim}[1]{\raisebox{0.5ex}{\scalebox{0.8}{$\displaystyle \lim_{#1}\;$}}}

\newcommand\Id{\hat{\mathds{1}}}

\newcommand{\SyO}{\hat{\mathcal{O}}}
\newcommand{\EnO}{\hat{\bar{\mathcal{O}}}}
\newcommand\bkouter[2]{\ket{#1}\bra{#2}}
\newcommand\Sys[1]{{#1}_{X }}
\newcommand\Env[1]{{#1}_{\bar{X} }}
\newcommand{\normor}[1]{\mathop{:}\nolimits #1 \mathop{:}\nolimits}
\title{\center{Mutual information as a measure of renormalizability}}

\author[a,1]{Brenden Bowen,\note{B. B. and A. F. contributed equally to this work.}}
\author[a,1]{Albert Farah,}
\author[a]{Spasen Chaykov,}
\author[a]{and Nishant Agarwal}

\affiliation[a]{Department of Physics and Applied Physics, University of Massachusetts, Lowell, MA 01854, USA}

\emailAdd{brenden\_bowen@student.uml.edu}
\emailAdd{albert\_farah@student.uml.edu}
\emailAdd{spasen\_chaykov@student.uml.edu}
\emailAdd{nishant\_agarwal@uml.edu}

\abstract{Renormalization is an essential technique in field-theoretic descriptions of natural phenomena, where the absence of a UV-complete description yields an abundance of divergent quantities. While the renormalization prescription has been thoroughly refined for equilibrium systems, consistently extending it to out-of-equilibrium systems is an active area of research. In this paper, we identify a mutual information-based measure of renormalizability that applies to quantum field theories both in and out of equilibrium. Specifically, we use mutual information to characterize correlations between infinitesimal shells in momentum space and show that the logarithmic derivative of mutual information with mode separation, at large mode separation, is a measure of renormalizability. We first consider Minkowski spacetime, where we introduce dynamics by performing an interaction quench, initializing the field in the free theory vacuum and then turning on the interaction. We show that the late-time mutual information relaxes to that for the interacting vacuum and the logarithmic derivative at large mode separation is negative for super-renormalizable theories, zero for renormalizable (marginal) theories, and positive for non-renormalizable theories. We then consider a conformally-coupled scalar field on the Poincar\'{e} patch of de Sitter spacetime, initializing the field in the Bunch-Davies vacuum in the asymptotic past. For different self-interactions and at any finite time, we find that the resulting mutual information has the same qualitative behavior as a function of mode separation, demonstrating that it can be used as a reliable indicator of renormalizability.}

\arxivnumber{2511.09625}

\begin{document}
\maketitle
\vspace{-0.75cm}
\section{Introduction}
\label{sec:intro}
The emergence of divergent integrals in loop corrections is one of the hallmarks of interacting quantum field theories (QFTs). Fundamentally, we attribute these divergences to the lack of a UV-complete theory and anticipate that the QFT will break down at energies above some finite scale. Nevertheless, we can often obtain a suitable low energy theory through a systematic absorption of such divergences into the parameters of the theory. In perturbation theory, this process is typically carried out in two steps. The first is \textit{regularization}, whereby we replace the divergent integrals with convergent, one-parameter generalizations, hence turning observables into finite quantities, except when the new parameter, called a regulator, is set to its value in the original divergent integrals. The second is \textit{renormalization}, in which we use the symmetries of the QFT to generate counterterms in the Lagrangian, and attempt to set these counterterms to cancel all instances of the regulator in observables. While regularization is generically possible, whether the theory is renormalizable or not is categorically dependent on the interaction Lagrangian \cite{Peskin:1995ev}. 

Pragmatically, one can assert that the only ``well-defined" QFTs are those with renormalizable interactions, for which counterterms can absorb all divergences. However, proper justification for this practice comes from the Wilsonian picture of renormalization where, guided by the goal of writing an effective field theory, one identifies the interactions that can be neglected at low energies; these \textit{irrelevant} interactions are precisely those that are non-renormalizable. Hence, whether a given field theory is renormalizable is intimately connected to how strongly momentum modes are correlated across scales \cite{Wilson:1971bg,Wilson:1973jj,Wilson:1983xri}. 

While the renormalization procedure has been thoroughly refined for equilibrium systems, a need to consistently extend this prescription to systems that necessitate the inclusion of finite-time effects remains. Additionally, there has been growing interest in applying techniques from quantum information theory to gain new perspectives on renormalization, renormalization group (RG) flow, and quantum criticality~\cite{Swingle:2010jz,Balasubramanian:2011wt,Beny:2012qh,Balasubramanian:2014bfa,Agon:2014uxa,Laflorencie:2015eck,Alves:2017fjk,Maity:2020qcg,FowlerIII:2020xgq,Fowler:2021oje,Paul:2022wgq,Cotler:2022fze,Zou:2022oxp,Costa:2022bvs,MartinsCosta:2022bjr,Kong:2025ksx}. Motivated by these developments, we identify a quantum information-based measure of renormalizability that naturally extends to out-of-equilibrium systems. 

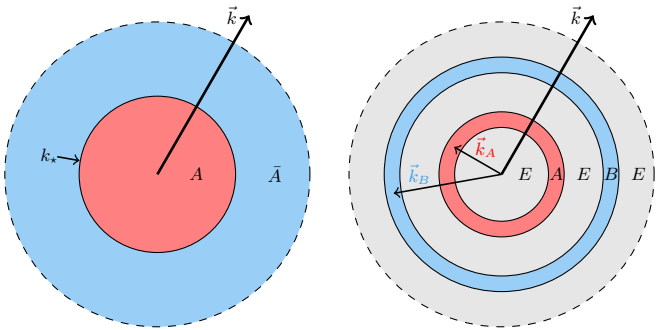
\begin{figure}[t!]
    \centering
    \begin{tikzpicture}[scale=0.65]
        \definecolor{Acolor}{RGB}{255,0,0};
		\definecolor{Bcolor}{RGB}{52,158,239};
        \draw[fill=Bcolor!50,dashed] (0,0) circle (3.9);
        \draw[fill=Acolor!50] (0,0) circle (2);
        \draw[->,line width=1] (0,0) -- (60:4.7);
        \node at (65:4.5){\scalebox{0.75}{$\vk$}};
        \node at (1,0){\scalebox{0.75}{$A$}};
        \node at (3,0){\scalebox{0.75}{$\bar{A}$}};
        \draw[->,line width=0.6] (170:2.6) -- (170:2.02);
        \node at (170:2.8){\scalebox{0.75}{$k_{\star}$}};
    \end{tikzpicture}
    \hspace{1cm}
    \begin{tikzpicture}[scale=0.65]
        \definecolor{Acolor}{RGB}{255,0,0};
		\definecolor{Bcolor}{RGB}{52,158,239};
        \draw[fill=gray!20,dashed] (0,0) circle (3.9);
        \draw[fill=Bcolor!50] (0,0) circle (3);
        \draw[fill=gray!20] (0,0) circle (2.6);
        \draw[fill=Acolor!50] (0,0) circle (1.6);
        \draw[fill=gray!20] (0,0) circle (1.2);
        \node at (0.6,0){\scalebox{0.75}{$E$}};
        \node at (2.1,0){\scalebox{0.75}{$E$}};
        \node at (3.5,0){\scalebox{0.75}{$E$}};
        \node at (0:1.4){\scalebox{0.75}{$A$}};
        \node at (0:2.8){\scalebox{0.75}{$B$}};
        \draw[->,line width=1] (0,0) -- (60:4.7);
        \node at (65:4.5){\scalebox{0.75}{$\vk$}};
        \draw[->,line width=0.6] (0,0) -- (150:1.4);
        \draw[->,line width=0.6] (0,0) -- (190:2.8);
        \node at (120:0.8){\textcolor{Acolor}{\scalebox{0.75}{$\vmodeone$}}};
        \node at (180:2.1){\textcolor{Bcolor}{\scalebox{0.75}{$\vmodetwo$}}};
    \end{tikzpicture}
    \caption{(Left) Bipartite partition of the Hilbert space, used to calculate the entanglement entropy of region $A$. (Right) Tripartite partition of the Hilbert space, used to calculate the mutual information between $A$ and $B$.
    }
    \label{fig:subspaces2D}
\end{figure}

One of the most fundamental quantities in quantum information theory is the von Neumann entropy, defined as
\begin{align}
    S = -\text{Tr} \hat{\rho} \ln \hat{\rho}
    \,.
\label{eq:vonNeumannEntropy}
\end{align}
for a given density operator $\hat{\rho}$. Among other interpretations, the von Neumann entropy quantifies how informationally far $ \hat{\rho} $ is from a pure state, and is identically zero when $ \hat{\rho} $ is a pure state. It also quantifies the {\it entanglement entropy} of a reduced system when $\hat{\rho}$ is obtained by performing a partial trace over environmental degrees of freedom on a joint density operator that is pure. This suggests that the von Neumann entropy can be used to characterize how low-energy (or IR) and high-energy (or UV) degrees of freedom are correlated in a QFT. As shown in the left panel of \cref{fig:subspaces2D}, we can partition the Hilbert space as $ \mathcal{H} = \mathcal{H}_{A} \otimes \mathcal{H}_{\bar{A}} $, where $\mathcal{H}_{A}$ and $\mathcal{H}_{\bar{A}}$ include all momentum modes below and above some scale $k_{\star}$, respectively. Tracing over the high-energy modes in $\bar{A}$ to obtain $\hat{\rho}_{A} = \text{Tr}_{\bar{A}} \hat{\rho}_{A\bar{A}}$ and using $\hat{\rho}_{A}$ in \cref{eq:vonNeumannEntropy}, we can then obtain a quantitative description of the entanglement between $A$ and $\bar{A}$ for both QFTs in equilibrium ~\cite{Balasubramanian:2011wt} and on cosmological spacetimes~\cite{Brahma:2020zpk,Brahma:2020rtx,Brahma:2022yxu,Brahma:2023hki}. While this procedure captures the information shared between the two regions as a whole, it does not distinguish between information shared with modes that are just outside the IR region and those that are infinitely outside it. Further, if we push $k_{\star} \to \infty$ in an attempt to capture the entanglement of the IR region with increasingly greater scales, then we find that $S_A$ always vanishes asymptotically for a pure $\hat{\rho}_{A\bar{A}}$ since $ \mathcal{H}_{A} $ ultimately subsumes $ \mathcal{H}_{\bar{A}} $~\cite{Balasubramanian:2011wt}. The entanglement entropy by itself is, therefore, not a reliable measure of correlations between IR and UV modes.

The essence of the ambiguity in using the entanglement entropy as a measure of correlations lies in the unavoidable shared boundary between $ \mathcal{H}_{A} $ and $ \mathcal{H}_{\bar{A}} $. An alternative measure of correlations between two systems $A$ and $B$ that interact with a common environment $E$, with the Hilbert space partitioned as $\mathcal{H} = \mathcal{H}_{A} \otimes \mathcal{H}_{B} \otimes \mathcal{H}_{E}$, is the entropy relative to the product state of the two reduced systems or the {\it quantum mutual information}, defined as
\begin{align}
\label{eq:MU1}
    I_{A:B} 
    = 
    \text{Tr}_{AB} \hat{\rho}_{AB} [\ln \hat{\rho}_{AB} -\ln \hat{\rho}_A \otimes \hat{\rho}_B]
    \,,
\end{align}
and written in terms of the von Neumann entropies as
\begin{align}
\label{eq:MU2}
    I_{A:B}
    =
    S_A + S_B - S_{A \cup B}
    \,.
\end{align}
As proposed in ref.~\cite{Balasubramanian:2011wt}, the mutual information between a mode in the IR region and another mode in the UV region is a more reliable measure of momentum-space correlations compared to the entanglement entropy. In this paper, we take the techniques of ref.~\cite{Balasubramanian:2011wt} further to calculate the mutual information between two infinitesimal shells $A$ and $B$ in momentum space, as shown in the right panel of \cref{fig:subspaces2D}. We show that the resulting mutual information as a function of mode separation can be used as a reliable indicator of renormalizability both in and out of equilibrium. It is worth noting that the fact that mutual information encompasses both entanglement and classical correlations~\cite{Groisman:2005dbo} is beneficial in developing it as a measure of renormalizability since {\it all} correlations, quantum and classical, should be finite after renormalization.

We first consider different self-interacting scalar field theories in Minkowski spacetime, where we introduce dynamics by performing an interaction quench, initializing the field in the free theory vacuum and then turning on the interaction. We calculate the time-evolved state to second order in perturbation theory and use \cref{eq:MU2} to obtain the mutual information between infinitesimal shells $ A $ and $ B $ defined by the modes $ \modeone = |\vmodeone| $ and $ \modetwo = |\vmodetwo| $ as shown in the right panel of \cref{fig:subspaces2D}. Since $k_B$ can be arbitrarily far from $k_A$, we can use the resulting mutual information to characterize  correlations as a function of mode separation, $r = k_B/k_A$. We find that the late-time mutual information relaxes to that for the interacting vacuum and the logarithmic derivative of mutual information with mode separation, at large mode separation, is negative for super-renormalizable theories, zero for renormalizable (marginal) theories, and positive for non-renormalizable theories. We then consider a dynamical spacetime, specifically a conformally-coupled scalar field on the Poincar\'{e} patch of de Sitter spacetime, initializing the field in the Bunch-Davies vacuum in the asymptotic past. For different self-interactions and at any finite time, we find that the resulting mutual information has the same qualitative behavior as a function of mode separation, demonstrating that it can be used as a reliable indicator of renormalizability both in and out of equilibrium.

The remainder of the paper is organized as follows. In \cref{sec:MImink}, we consider a self-interacting scalar field theory in Minkowski spacetime. We first write the normal ordered Hamiltonian such that all bubble contributions are subtracted out, with details given in \cref{app:NormalOrdering}. We then review the perturbative entanglement entropy calculation of ref.~\cite{Balasubramanian:2011wt}, further introducing dynamics following an interaction quench, in \cref{sec:MomentumSpaceEE}. We relegate details on diagonalizing the density operator to \cref{app:entropycalc} and on writing the entanglement entropy in terms of propagators to \cref{app:EEndDerivation}. In \cref{sec:MImomentumshells}, we calculate the mutual information density between two infinitesimal shells in momentum space, with details given in \cref{app:ndderivation}. We then calculate the resulting mutual information for specific interactions and demonstrate that it can be used as a measure of renormalizability, with details on computing certain integrals relegated to \cref{app:calcspecs}. In \cref{sec:TwoModeMI}, we derive a scaling relation for the mutual information between two modes in the limit of large mode separation and show that it agrees with the results of the previous subsection. In \cref{sec:MIdS}, we generalize our calculations to a conformally-coupled scalar field on the Poincar\'{e} patch of de Sitter spacetime and demonstrate similar qualitative behavior of the mutual information as a function of mode separation. We end with a discussion of our results including potential connections to RG flow in \cref{sec:disc}.

\section{Mutual information in Minkowski spacetime}
\label{sec:MImink}

We first consider a Klein-Gordon field, $ \phi(t, \vx) $, with mass $ m $, in $d+1$D Minkowski spacetime. We allow the field to evolve freely from the infinite past until the initial time $ t_{0} $ when we perturb the system via a $ \phi^{n} $ interaction of strength $ \lambda $. The full action after $t_0$ is then given by
\begin{align}
    S[\phi]
    =
    -
    \int \d^{d+1} x 
    \left[
        \frac{1}{2} (\partial \phi)^2 
        + 
        \frac{1}{2} m^2 \phi^2 
        + 
        \frac{\lambda}{n!} \phi^n
    \right],
\end{align}
where $ (\partial \phi)^2 = \eta^{\mu \nu} \del_{\mu} \phi \del_{\nu} \phi $ and $ \eta^{\mu}_{\hphantom{\mu} \nu} = \text{diag}(-1, 1, \dots, 1) $ is the Minkowski metric. Assuming weak coupling, and working in units where $ \hbar = c = 1 $, we quantize the field by imposing the canonical commutation relations $ [\hphi(\vx),\hphi(\vy)] = [\hpi(\vx),\hpi(\vy)] = 0 $ and $ [\hphi(\vx),\hpi(\vy)] = i \delta^{d}(\vx-\vy) $. We next write the interaction picture field operator in terms of its spatial Fourier modes as
\begin{align}
    \hphi (t, \vx) 
    = 
    \int_{\vk \inn \mathbb{R}^{d}}
    \hphi_{\vk}(t) 
    e^{i \vk \cdot \vx}
    \,,
\end{align}
where we introduced the convenient notation $ \int_{\vk \inn \mathbb{R}^{d}} \equiv \int_{\mathbb{R}^{d}} \frac{\d^d k}{(2\pi)^d} $ and excluded an interaction picture label since we work exclusively in the interaction picture. Defining the Schr\"odinger picture ladder operators $ \hc_{\vk} $ and $ \hcdagg_{\vk} $, which satisfy the commutation relations $ [\hc_{\vk},\hc_{\vk'}] = [\hcdagg_{\vk},\hcdagg_{\vk'}] = 0 $ and $[\hc_{\vk},\hcdagg_{\vk'}] = (2 \pi)^{d} \delta^{d}(\vk-\vk') $, we can then write the field operator as
\begin{align}
    \hat{\phi}_{\vk} (t)
    =
    f_{k}^{>} (t) \hc_{\vk}
    +
    f_{k}^{<} (t) \hcdagg_{-\vk}
    \,,
\end{align}
where the free theory mode functions are given by
\begin{align}
    f_{k}^{>} (t)
    =
    \frac{1}{\sqrt{2 \omega_k}} e^{-i \omega_k (t-t_0)}
    =
    f_{k}^{<*} (t),
\label{eq:ModeFunc}
\end{align}
where $ \omega_k \equiv \sqrt{k^2 + m^2}$. The Hamiltonian now takes the form
\begin{align}
\label{eq:phiNHamiltonian}
    \hat{H} 
= 
    \int_{\vk \inn \mathbb{R}^{d}}  
    \omega_{k}
    \normor{
        \hcdagg_{\vk}
        \hc_{\vk}
    }
    +
    \frac{\lambda}{n!}
    \int \d^{d} x 
    \normor{
         \hat{\phi}^{n}
    }
    \,,
\end{align}
where we use $ \normor{\bullet} $ to denote that we have normal ordered the Hamiltonian so that its expectation in the initial state, which we take to be the vacuum, is zero. For our calculations below, it is useful to express a normal ordered product of field operators in terms of correlation functions, explicitly subtracting out bubble contributions. We obtain such an expression for any Gaussian state in \cref{app:NormalOrdering}, and show that it agrees with the standard procedure of ``moving all annihilation operators to the right'' for the vacuum case.

\subsection{Entanglement entropy in momentum space}
\label{sec:MomentumSpaceEE}

To obtain the mutual information between modes in regions $ A $ and $ B $, as shown in the right panel of \cref{fig:subspaces2D}, we must first calculate the entanglement entropies $ S_{A} $, $ S_{B} $, and $ S_{A \cup B} $. Since the same procedure can be used for all three entanglement entropies, we first calculate the entanglement entropy $S_X$ between any momentum space region $X$ and its complement $ \bar{X} $ and then choose $ X \in \{A, B, A \cup B\} $. We start by defining a \textit{system} field operator $ \hphi_{X}(t,\vx) $ and an \textit{environment} field operator $ \hphi_{\bar{X}}(t,\vx) $ such that their $\vk$ modes are in the regions $ X $ and $ \bar{X} $, respectively,
\begin{align}
    \hat{\phi}_{X}(t,\vx) 
    = 
    \int_{\vk \inn X} 
    \hat{\phi}_{\vk}(t) 
    e^{i \vk \cdot \vx}
    \quad\text{and}\quad
    \hat{\phi}_{\bar{X}}(t,\vx) 
    = 
    \int_{\vk \inn \bar{X}} 
    \hat{\phi}_{\vk}(t) 
    e^{i \vk \cdot \vx}
    \,.
\end{align}
Note that $  \hat{\phi}(t,\vx)  = \hat{\phi}_{X}(t,\vx)  + \hat{\phi}_{\bar{X}}(t,\vx)  $ since $ X \cup \bar{X} = \mathbb{R}^{d} $. We can now write the interaction term in the Hamiltonian in \cref{eq:phiNHamiltonian} as a sum of composite operators that act explicitly on the system or the environment. In the interaction picture, we have
\begin{align}
    \hat{H}_{\text{I}} (t)
    &=
    \lambda
    \sum_{\alpha=0}^{n}
        \int \text{d}^d x
        \SyO_{n - \alpha} (t, \vx)
        \EnO_{\alpha} (t, \vx)
    \,,
    \label{eq:IntHam}
\end{align}
where the system and environment operators are
\begin{align}
    \SyO_{n - \alpha} (t, \vx) 
    &\equiv 
    \frac{
        1
    }{
        \left(
            n
            -
            \alpha
        \right)!
    }
    \normor{ \hat{\phi}_{X}^{n-\alpha} (t, \vx) } \quad \text{and}
    \label{eq:SysOp}
\\
    \EnO_{\alpha} (t, \vx) 
    &\equiv 
    \frac{
        1
    }{
        \alpha!
    }
    \normor{\hat{\phi}_{\bar{X}}^\alpha (t, \vx)}
    \,,
    \label{eq:EnvOp}
\end{align}
respectively. 

We next assume that, in the distant past, the system is in the vacuum of the free theory, which we write in terms of system and environment modes as
\begin{align}
    \ket{0}
    =
    \Big(\mathlarger{\textstyle \bigotimes_{\vk \inn X} } \ket{0_{\vk}} \Big)
    \hspace{-2pt}
    \otimes
    \hspace{-2pt}
    \Big(\mathlarger{\textstyle \bigotimes_{\vk \inn \bar{X}} } \ket{0_{\vk}} \Big)
    \equiv
    \ket{0_{X}, 0_{\bar{X}}}
    \,,
\end{align}
and note that the normal ordering in \cref{eq:SysOp,eq:EnvOp} automatically implies that \\ $ \langle \Sys{0}| \SyO_{\alpha} | \Sys{0} \rangle = 0 $ and $ \langle \Env{0}| \EnO_{\alpha} | \Env{0} \rangle = 0 $. Since the free theory vacuum is not an eigenstate of the quenched Hamiltonian, $ \ket{0} $ undergoes a non-trivial time evolution beginning at $ t_{0} $. In the interaction picture, this time-evolved state is
\begin{align}
\label{eq:QunchdSt}
    \ket{\Omega (t)}
=
    \hat{U}(t,t_0)
    \ket{0}
    ,
\end{align}
where the interaction picture time evolution operator is defined as
\begin{align}
    \hat{U}(t,t_0)
    =
    T e^{-i \int_{t_1}  \hat{H}_{\text{I}}(t_1)}
    \,,
\end{align}
with $T$ indicating time-ordering and $ \int_{t_1} \equiv \int_{t_0}^t \text{d} t_1$. From this, we construct the time-dependent density operator of the composite system and perform a partial trace over the $\bar{X}$ degrees of freedom to obtain the reduced density operator,
\begin{align}
    \hat{\rho}_X (t)
=
    \text{Tr}_{\bar{X}}
    |\Omega (t)\rangle \langle\Omega (t)|
    \,.
\end{align}
In principle, using this in \cref{eq:vonNeumannEntropy} yields the entanglement entropy between $ X $ and $ \bar{X} $. However, since the $ \lambda \phi^{n} $ interactions we consider are not exactly solvable, the calculation of entanglement entropy requires the use of perturbation theory. 

In \cref{app:entropycalc}, we review the techniques of ref.~\cite{Balasubramanian:2011wt} to obtain the eigenvalues of the reduced density operator to second order in $ \lambda $ and use the result to write the entanglement entropy at leading order. We find that
\begin{align}
    S_{X} (t)
    &=
    -
    \lambda^2 \ln(\bar{\lambda}^2)
    \sum_{\alpha, \beta}
    \int_{x_1, x_2}
    \braket{\Env{0}|
        \EnO_{\alpha}(x_1)
        \EnO_{\beta}(x_2)
    |\Env{0}}_c
    \braket{\Sys{0}|
        \SyO_{n-\alpha}(x_1)
        \SyO_{n-\beta}(x_2)
    |\Sys{0}}_c
    ,
    \label{eq:SvN}
\end{align} 
where $x_i \equiv (t_i, \vx_{i})$ and $\int_{x_i} \equiv \int_{t_0}^{t} \d t_i \int \d^d x_i$, $ \bar{\lambda} $ is a dimensionless coupling constant defined in \cref{app:entropycalc}, and $\braket{0|\SyO_1\SyO_2|0}_c = \braket{0|\SyO_1\SyO_2|0} - \braket{0|\SyO_1|0} \braket{0|\SyO_2|0}$ is the connected vacuum correlator. Note that since $ \SyO $ and $ \EnO $ are normal ordered products of field operators, their one-point functions vanish and $ \braket{0|\SyO_1\SyO_2|0}_c = \braket{0|\SyO_1\SyO_2|0} $. However, we keep the subscript $ c $ for generality since, as shown in \cref{app:entropycalc}, connected correlators appear naturally in the computation of entanglement entropy for a field theory Hamiltonian of the form in \cref{eq:IntHam}.

In \cref{app:EEndDerivation}, we expand the entanglement entropy in terms of field operators using \cref{eq:SysOp,eq:EnvOp}, and employ Wick's contractions to reduce the correlation functions in \cref{eq:SvN} to products of propagators. We find that only diagrams that are fully connected loops between $ x_{1} $ and $ x_{2} $ contribute to the entanglement entropy and, since there are no external lines, the integrals over $ \vx_{1} $ and $ \vx_{2} $ produce identical momentum conserving delta functions when transforming to momentum space (essentially due to the extensivity of the self-energy). We, therefore, work with the entanglement entropy {\it density} by dividing out a volume factor of $ L^{d} $, obtaining
\begin{align}
    \frac{S_{X} (t)}{L^d}
    =
    &-\frac{\lambda^2 \ln \bar{\lambda}^2}{n!} 
    \int_{t_1, t_2}
    \Bigg[
    \prod_{i=1}^{n}
        \int_{\vk_{i} \in \mathbb{R}^d}
    -
    \prod_{i=1}^{n}
        \int_{\vk_{i} \in X}
    -
    \prod_{i=1}^{n}
        \int_{\vk_{i} \in \bar{X}}
    \Bigg] 
\nn \\
& \quad \times
    (2\pi)^{d} 
    \delta^d 
    \Big(
        \mathlarger{\textstyle \sum_{m=1}^{n}} \vk_{m}
    \Big)
    \prod_{j=1}^{n}
        G_{k_j}^{>} (t_1, t_2)
    \,,
    \label{eq:infvolEE}
\end{align}
where we defined the propagator $G^>_{k_i} (t_1, t_2) \equiv f_{k_i}^{>} (t_1) f_{k_i}^{<} (t_2)$. 

\subsection{Mutual information between momentum shells}
\label{sec:MImomentumshells}

We now consider the mutual information between two infinitesimal shells $A$ and $B$ around the momentum modes $ \modeone $ and $ \modetwo $, as depicted in \cref{fig:subspaces2D}. Since we found that the entanglement entropy must be defined as a density, we similarly work with the mutual information density, $ I_{A:B}/L^{d} $. In \cref{app:ndderivation}, we use \cref{eq:infvolEE} in \cref{eq:MU2} to write the mutual information density as
\begin{align}
    \frac{I_{A:B} (t)}{L^d}
&=
    -\frac{\lambda^2 \ln \bar{\lambda}^2}{n!} 
    \int_{t_1, t_2}
    \int_{A:B}
    \prod_{j=1}^{n}
    G_{k_j}^{>} (t_1, t_2)    
    \,,
\label{eq:FullMI}
\end{align}
where we defined the shorthand
\begin{align}
    \int_{A:B}
\equiv
    &\left[
        \prod_{i=1}^{n}
            \int_{\vk_{i} \inn \mathbb{R}^d} 
        -
        \prod_{i=1}^{n}
            \int_{\vk_{i} \inn A} 
        -
        \prod_{i=1}^{n}
            \int_{\vk_{i} \inn B} 
        +
        \prod_{i=1}^{n}
            \int_{\vk_{i} \inn A \cup B} 
        -
        \prod_{i=1}^{n}
            \int_{\vk_{i} \inn \bar{A}} 
        -
        \prod_{i=1}^{n}
            \int_{\vk_{i} \inn \bar{B}} 
        +
        \prod_{i=1}^{n}
            \int_{\vk_{i} \inn E} 
    \right]
\nn \\
&\quad \times
    (2 \pi)^{d}
    \delta^{d}
    \left(
        \mathlarger{\textstyle \sum_{m=1}^{n}} \vk_{m}
    \right)
    \,.
\end{align}
We now assume that $A$ and $B$ are disjoint regions, so that $ \bar{A}  =  B \cup E $, $ \bar{B}  =  A \cup E $, and $ \int_{\vk \inn A \cup B}  =  \int_{\vk \inn A} + \int_{\vk \inn B} $. Then leveraging that $ A $ and $ B $ are of infinitesimal thickness $\delta k$, we perform the radial integral in $ \int_{\vk_{i} \in A} $ by evaluating the integrand at $ k_{i} = \modeone $ and multiplying by $ \delta k $, and similarly for $ B $, to expand \cref{eq:FullMI} to leading order in $ \delta k $. Finally, defining the ratio $ r = \modetwo / \modeone $ and denoting the mutual information density $  I_{\modeone:r\modeone} (t) / L^d \equiv \mathcal{I}_{r}(t) $, we find in \cref{app:ndderivation} that $\mathcal{I}_{r}(t)$ can be written as
\begin{align}
    \mathcal{I}_{r}(t)
=
    &-
    \frac{
        \lambda^2 \ln \bar{\lambda}^2
        \delta k^{2}
        \modeone^{2d-2}
        r^{d-1}
    }{
        (2 \pi)^{2d}
        (n-2)!
    }
    \int_{t_1, t_2}
    G_{\modeone}^{>}(t_1, t_2)
    G_{r\modeone}^{>}(t_1, t_2)
\nn \\
&\quad \times
    \int_{\Omega_{1}, \Omega_{2}}
    \left[
        \prod_{i=3}^{n}
            \int_{\vk_{i} \inn \mathbb{R}^d} 
            G_{k_{i}}^{>} (t_1, t_2)
    \right]
    (2 \pi)^{d}
    \delta^{d}
    \big(
        \vec{K}_{r}^{(n)}
    \big)
    \,,
\label{eq:MIr}
\end{align}
where $ \vec{K}_{r}^{(n)} \equiv \modeone(\hat{k}_{1} + r \hat{k}_{2}) + \sum_{m=3}^{n} \vk_{m} $, with $ \hat{k}_{1} $ and $ \hat{k}_{2} $ being unit vectors, and $ \int_{\Omega_{i}} $ is an integral over the solid angle of dimension $d-1$, corresponding to $ \vk_{i} $.  Note, in particular, that $ \mathcal{I}_{1} $, for $r = 1$, is \textit{not} equivalent to the second order $ \delta k $ expansion of $ S_{A}/L^{d} $, as one might guess from \cref{eq:MU2}, since we have explicitly used $ A \cap B = \emptyset $ when performing the expansion in $ \delta k $. However, $ \mathcal{I}_{1} $ is still a finite quantity, and will be more convenient than the $ \delta k $ expansion of $ S_{A}/L^{d} $ to normalize $ \mathcal{I}_{r} $ in our resuts below, since the coefficients of $ \mathcal{I}_{r} $ and $ \mathcal{I}_{1} $ exactly cancel, reducing the number of parameters and allowing for more general conclusions.

An important limiting case of \cref{eq:MIr} is the late-time limit, that can be obtained analytically as follows. Using \cref{eq:ModeFunc} in $G^>_{k} (t_1, t_2) = f_{k}^{>} (t_1) f_{k}^{<} (t_2)$, we compute the late-time limit of the time integrals in \cref{eq:MIr} as
\begin{align}
    \int_{0}^{\infty}
    \d t_{1}
    \int_{0}^{\infty}
    \d t_{2}
    e^{-i(\omega_{\modeone} + \omega_{r\modeone} + {\scriptstyle \sum_{i=3}^{n} \omega_{k_{i}}})(t_{1} - t_{2})}
=
    \frac{
        1
    }{
        (
        \omega_{\modeone} 
            + 
            \omega_{r\modeone} 
            + 
            \sum_{i=3}^{n} \omega_{k_{i}}
        )^{2}
    }
    \, ,
\end{align}
where we used $ \int_{0}^{\infty} \d t e^{\pm i \omega t} = \pi \delta(\omega) \pm i P(1/\omega) $ to evaluate the integrals, imposed $ \omega_{\modeone}  +  \omega_{r\modeone}  +  \sum_{i=3}^{n} \omega_{k_{i}} \neq 0 $ to drop the delta function contributions, and removed the superfluous $P$ from the principal value piece. The late-time limit of the mutual information density is then given by
\begin{align}
    \mathcal{I}_{r}(\infty)
=
    &-
    \frac{
        \lambda^2 \ln \bar{\lambda}^2
        \delta k^{2}
        \modeone^{2d-2}
        r^{d-1}
    }{
        2^{n}
        (2 \pi)^{2d}
        (n-2)!
    }
    \int_{\Omega_{1}, \Omega_{2}}
    \left[
        \prod_{i=3}^{n}
            \int_{\vk_{i} \inn \mathbb{R}^d} 
    \right]
\nn \\
&\quad \times
    \frac{
        (2 \pi)^{d}
        \delta^{d}
        \big(
            \vec{K}_{r}^{(n)}
        \big)
    }{
        \omega_{\modeone} 
        \omega_{r\modeone} 
        \omega_{k_{3}}
        \cdots
        \omega_{k_{n}}
        (
            \omega_{\modeone} 
            + 
            \omega_{r\modeone} 
            + 
            \sum_{i=3}^{n} \omega_{k_{i}}
        )^{2}
    }
    \,,
\label{eq:MIrTI}
\end{align}
which matches the result that one would obtain for the ground state of the interacting theory using time-independent perturbation theory, as in ref.~\cite{Balasubramanian:2011wt}. We thus refer to $ \mathcal{I}_{r}(\infty) $ as the time-independent mutual information density.

We now calculate the mutual information density for the specific cases $ n =  3 $, $ 4 $, and $ 6 $, and for different $d$ in sections\ \ref{sec:MImink3} to \ref{sec:MImink6} below, to demonstrate that it can be used as a measure of renormalizability.

\begin{figure*}[!t]
\centering
    \includegraphics[scale=0.5]{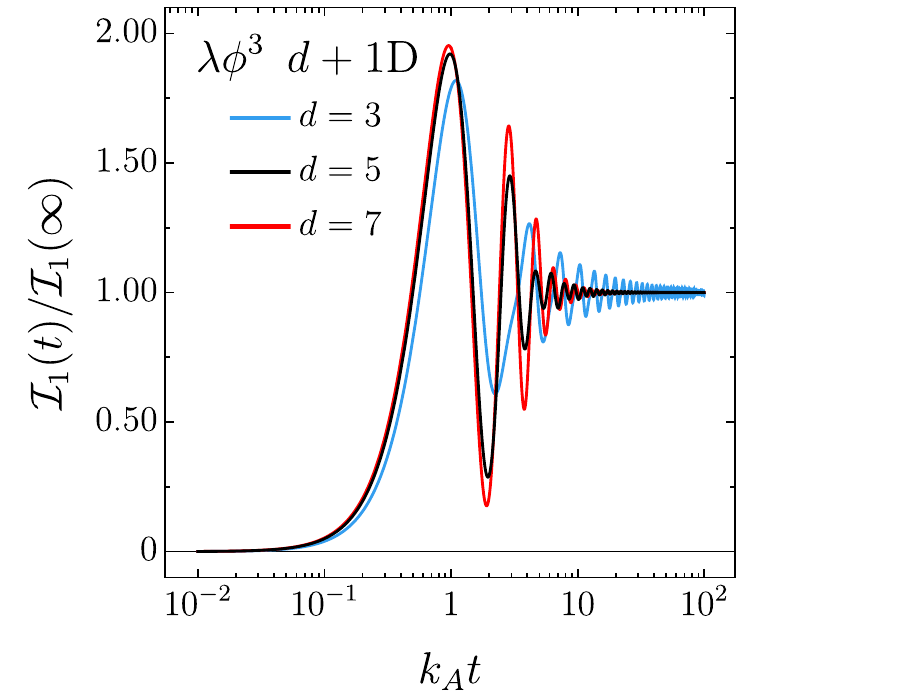}
    \includegraphics[scale=0.5]{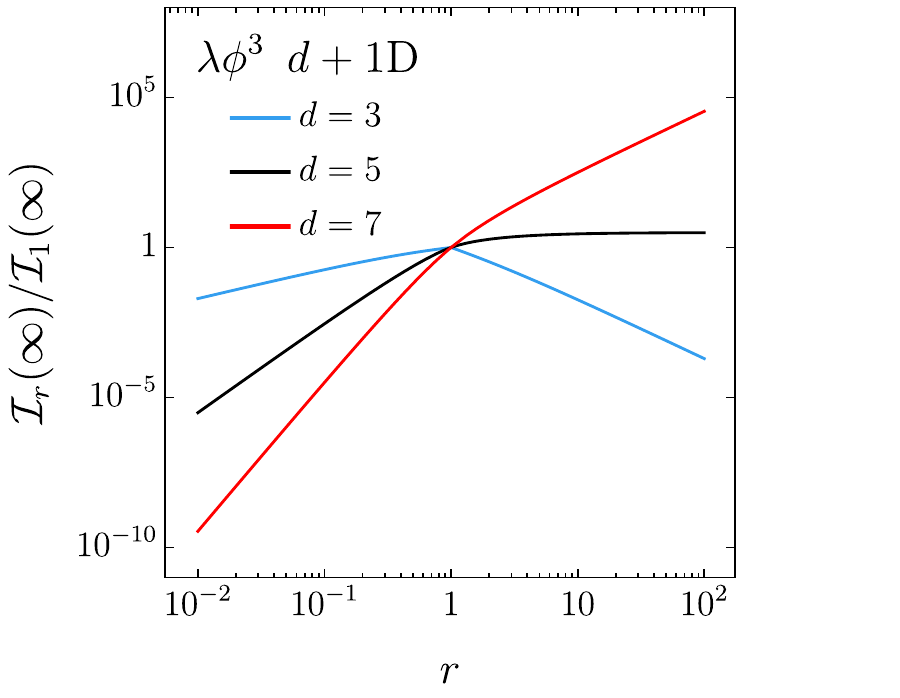}
    \caption{(Left) Time-dependent mutual information density as a function of time for a massless $\lambda \phi^3$ theory in $d = 3$, $5$, and $7$ spatial dimensions for the $r = 1$ case, normalized by the time-independent result. We find the same qualitative behavior for any $r$. (Right) Time-independent mutual information density as a function of mode separation for the same theory, normalized to unity at $r = 1$.}
    \label{fig:TDTIminkowskiphi3}
\end{figure*}

\subsubsection{\texorpdfstring{$\lambda \phi^3$ in $d+1$D}{λϕ3 in d + 1D}}
\label{sec:MImink3}

We first compute the time-dependent and time-independent mutual information density in \cref{eq:MIr,eq:MIrTI} for a massless $ \lambda \phi^{3} $ theory in $d = 3$, $5$, and $7$ spatial dimensions. In this case, there is only one mode other than $ \modeone $ and $ \modetwo $ and, therefore, after performing the integral over the momentum-conserving delta function, we are only left with the angular integrals over the regions $ A $ and $ B $, that we compute in \cref{app:phi3cal}. As shown in the left panel of \cref{fig:TDTIminkowskiphi3}, the resulting time-dependent mutual information for the $r = 1$ case oscillates around the time-independent result for times greater than $ \modeone^{-1} $ and converges to it at late times, as expected, converging faster for higher $d$. We find the same qualitative behavior for any mode separation $r$, and thus restrict to the time-independent mutual information density in the right panel of \cref{fig:TDTIminkowskiphi3}. As shown there, for $d = 3$, where the $\lambda \phi^3$ theory is super-renormalizable, the mutual information density, normalized by the $r = 1$ result, decays to zero as $ r \to \infty $. Further, we find that the logarithmic derivative of the mutual information with mode separation, at large mode separation, is $\vphantom{\Big(} \Lim{r \to \infty} \frac{\d \ln \mathcal{I}_{r} (\infty)}{\d \ln r} = -2$. For $d = 5$, where the $\lambda \phi^3$ theory is renormalizable, on the other hand, the mutual information density saturates as $ r \to \infty $ and $\vphantom{\Big(} \Lim{r \to \infty} \frac{\d \ln \mathcal{I}_{r} (\infty)}{\d \ln r} = 0$. Lastly, for $d = 7$, where the $\lambda \phi^3$ theory is non-renormalizable, the mutual information density grows indefinitely as $ r \to \infty $ and $\vphantom{\Big(} \Lim{r \to \infty} \frac{\d \ln \mathcal{I}_{r} (\infty)}{\d \ln r} = 2$.

\begin{figure*}[!t]
\centering
    \includegraphics[scale=0.5]{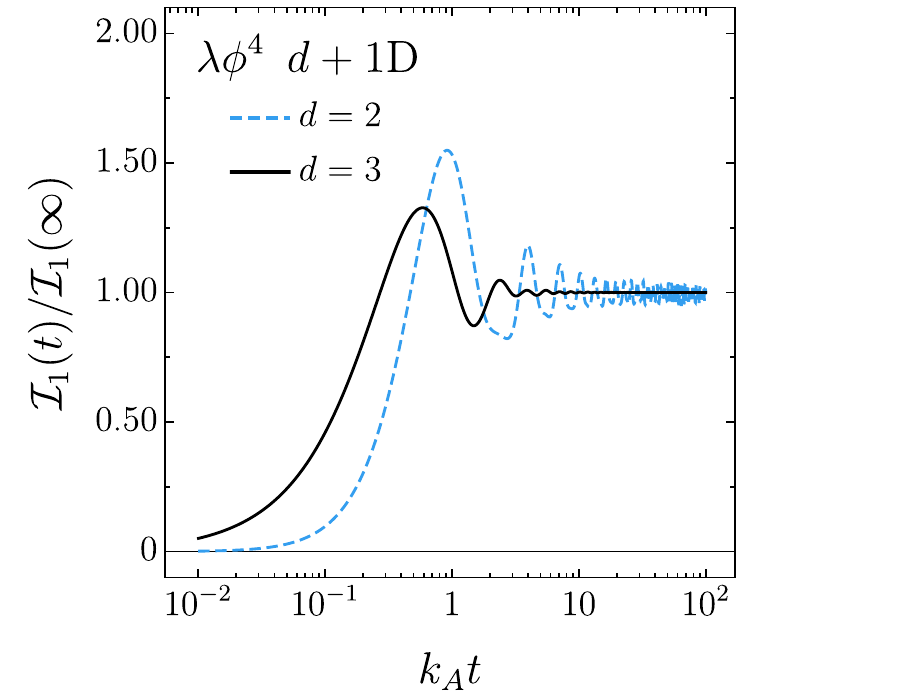}
    \includegraphics[scale=0.5]{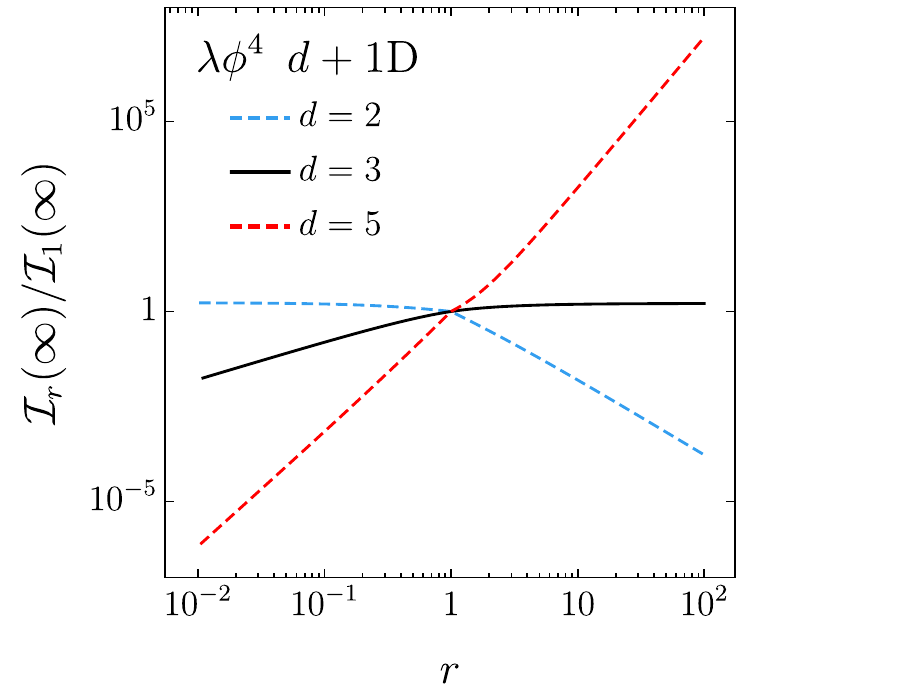}
    \caption{(Left) Time-dependent mutual information density as a function of time for a massless $\lambda \phi^4$ theory in $d = 2$ and $3$ spatial dimensions for the $r = 1$ case, normalized by the time-independent result. Solid and dashed lines indicate analytic and numerical results, respectively. We find the same qualitative behavior for any $r$. (Right) Time-independent mutual information density as a function of mode separation for the same theory, normalized to unity at $r = 1$. For the additional $\lambda \phi^4$ in $5+1$D case, we regularized the integrals using dimensional regularization with a regularization parameter $ \xi = 10^{-3}$.}
    \label{fig:TDTIminkowskiphi4}
\end{figure*}

\subsubsection{\texorpdfstring{$\lambda \phi^4$ in $d+1$D}{λϕ4 in d + 1D}}
\label{sec:MImink4}

We next consider a massless $ \lambda \phi^{4} $ theory in $d = 2$, $3$, and $5$ spatial dimensions. In this case, there are two modes other than $ \modeone $ and $ \modetwo $ and, therefore, after evaluating the momentum-conserving delta function, we are left with one integral over $ \mathbb{R}^{d} $ in addition to the angular integrals over the regions $ A $ and $ B $, that we compute in \cref{app:phi4cal}. By simple power counting, we find that the integrals are convergent in $ d = 2$ and $3$, where $ \lambda \phi^{4} $ is renormalizable, and divergent in $ d = 5 $, where $\lambda \phi^{4} $ is non-renormalizable. We thus regulate the $d = 5$ integral using dimensional regularization and further use numerical techniques to compute the $d = 2$ and $5$ integrals. As shown in the left panel of \cref{fig:TDTIminkowskiphi4} for $d = 2$ and $3$, the resulting time-dependent mutual information for the $r = 1$ case again converges to the time-independent result at late times. For $d = 5$, the regulated time-dependent expression with two time integrals is not amenable to numerical integration and so we do not include this result, though we expect it to show the same relaxation toward the time-independent result at late times. We also find the same qualitative behavior for any mode separation $r$, and thus again restrict to the time-independent mutual information density in the right panel of \cref{fig:TDTIminkowskiphi4}. As shown there, for $d = 2$, where the $\lambda \phi^4$ theory is super-renormalizable, the mutual information density, normalized by the $r = 1$ result, decays to zero as $ r \to \infty $ and  $\vphantom{\Big(} \Lim{r \to \infty} \frac{\d \ln \mathcal{I}_{r} (\infty)}{\d \ln r} = -2$. For $d = 3$, where the $\lambda \phi^4$ theory is renormalizable, on the other hand, the mutual information density saturates as $ r \to \infty $ and $\vphantom{\Big(} \Lim{r \to \infty} \frac{\d \ln \mathcal{I}_{r} (\infty)}{\d \ln r} = 0$. Lastly, for $d = 5$, where the $\lambda \phi^4$ theory is non-renormalizable, the mutual information density grows indefinitely as $ r \to \infty $ and $\vphantom{\Big(} \Lim{r \to \infty} \frac{\d \ln \mathcal{I}_{r} (\infty)}{\d \ln r} = 4$.

\subsubsection{\texorpdfstring{$\lambda \phi^6$ in $2+1$D}{λϕ6 in 2 + 1D}}
\label{sec:MImink6}

\begin{figure}[!t]
    \centering    
    \includegraphics[scale=0.5]{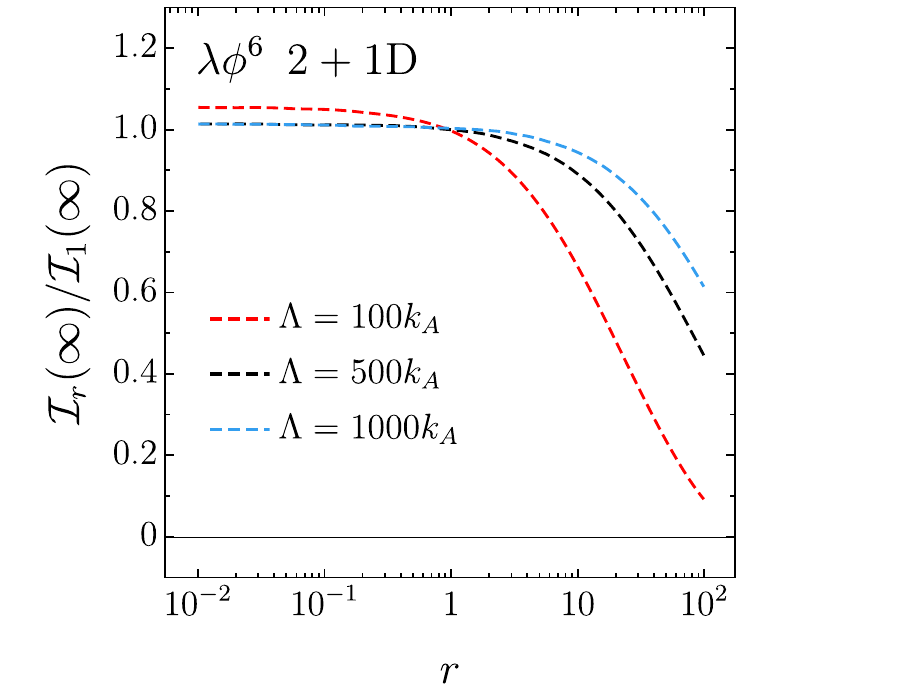}
    \caption{Time-independent mutual information density as a function of mode separation for a massless $\lambda \phi^6$ theory in $2+1$D, normalized to unity at $ r = 1 $. All integrals were computed numerically for different choices of UV cutoffs $\Lambda$ as shown.
    }
    \label{fig:minkowskiphi6}
\end{figure}

Lastly, we consider a massless $ \lambda \phi^{6} $ theory in $d = 2$ spatial dimensions. In this case, there are four modes other than $ \modeone $ and $ \modetwo $ and, therefore, after evaluating the momentum-conserving delta function, we are left with three integrals over $ \mathbb{R}^{d} $ in addition to the angular integrals over the regions $ A $ and $ B $. Unlike the previous renormalizable cases, these remaining momentum integrals (in the mutual information) are not power counting-convergent~\cite{Balasubramanian:2011wt}. To understand this behavior better, we regulate the integrals using a UV cutoff $ \Lambda $ and study the time-independent mutual information density in \cref{eq:MIrTI} as a function of both $r$ and $ \Lambda $, again using numerical techniques to compute the integrals. As shown in \cref{fig:minkowskiphi6}, the mutual information is uniform for $ r \ll \Lambda $, with $ \mathcal{I}_{r}(\infty) \approx \mathcal{I}_{1}(\infty) $, and begins to decay to zero as $ r $ approaches $ \Lambda $. Since this uniform behavior extends to larger $ r $ as we increase $ \Lambda $, we deduce that $ \mathcal{I}_{r}(\infty) \approx \mathcal{I}_{1}(\infty) $ for \textit{all} $ r $ in the limit $ \Lambda \to \infty $. It is also worth noting that since $ \mathcal{I}_{1}(\infty) $ itself diverges as $ \Lambda \to \infty $, $ \mathcal{I}_{r}(\infty) $ must go to infinity uniformly for all $r$, so that $\vphantom{\Big(} \Lim{r \to \infty} \frac{\d \ln \mathcal{I}_{r} (\infty)}{\d \ln r} = 0$, consistent with the previous renormalizable cases for $ \lambda \phi^{3} $ and $ \lambda \phi^{4} $.

We summarize the conclusions of this subsection in \cref{tab:summarytable}.

\subsection{Mutual information between two modes}
\label{sec:TwoModeMI}

In this subsection, we derive a scaling relation for the time-independent mutual information between two modes for $ r \gg 1 $, to gain an analytical understanding of the results obtained in the previous subsection. Consider the general expression for the time-independent mutual information between two modes $ \vmodeone $ and $ \vmodetwo $ with magnitudes $ \modeone $ and $ r \modeone $ for a massless $ \lambda \phi^{n} $ theory in $d+1$D,
\begin{align}
    I_{\vmodeone:\vmodetwo}
&\sim 
    \int_{1 - \frac{\delta k}{2 \modeone}}^{1 + \frac{\delta k}{2 \modeone}}
    \d z_{1}
    \int_{r - \frac{\delta k}{2 \modeone}}^{{r + \frac{\delta k}{2 \modeone}}}
    \d z_{2}
    \int_{\delta \Omega_{1}, \delta \Omega_{2}}  
    \hspace{-3pt}
    \left[
        \prod_{j=3}^{n}
        \int_{\vec{z}_{j} \inn \mathbb{R}^{d}}  
    \right]
    \frac{
        z_{1}^{d-1}
        z_{2}^{d-1}
        \delta^{d} 
        \big(
            \vec{z}_{1} 
            + 
            \vec{z}_{2} 
            + 
            \vec{Z}
            +
            \vec{z}_{n}
        \big)
    }{
        z_{1}
        z_{2}
        \cdots
        z_{n}
        \big(
            z_{1} 
            + 
            z_{2} 
            + 
            ||\vec{Z}||_{1}
            +
            z_{n}
        \big)^2
    }
    ,
\end{align}
where $ \sim $ indicates that we have dropped all $r$-independent factors, $ \int_{\delta \Omega_{i}} $ is an integral over the infinitesimal solid angle of dimension $d-1$, and we have defined dimensionless integration variables $ \vec{z}_{i} $ as $ \vk_{i} = \modeone \vec{z}_{i} $, as well as the shorthand $ \vec{Z} = \sum_{i=3}^{n-1} \vec{z}_{i} $ and $ ||\vec{Z}||_{1} = \sum_{i=3}^{n-1} z_{i} $. We can evaluate the $ \vec{z}_{n} $ integral over the delta function and then use the fact that the resulting integrand is approximately constant over the limits of the $ z_{1} $ and $ z_{2} $ integrals to evaluate them as well. Dropping $ \delta k^{2} $ and $ \delta \Omega^{2} $, we find that
%
\begin{table}[t!]
\centering
\begin{tabular}{|c|c|c|c|}
    \hline
        $n$
    &
        $d$
    & 
        $\vphantom{\Big(} \Lim{r \to \infty} \frac{\d \ln \mathcal{I}_{r} (\infty)}{\d \ln r}$
    &
        Renormalizability
    \\
    \hline
    \hline
      & 3 &  $-2\hphantom{-}$ & Super-renormalizable \\
    3 & 5 &  $0$ & Renormalizable \\
      & 7 &  $2$ & Non-renormalizable \\
    \hline
      & 2 &  $-2\hphantom{-}$ & Super-renormalizable \\
    4 & 3 &  $0$ & Renormalizable \\
      & 5 &  $4$ & Non-renormalizable \\
    \hline
    6 & 2 &  $0$ & Renormalizable  \\
    \hline
\end{tabular}
\caption{Summary of the logarithmic derivative of mutual information with mode separation, at large mode separation, for different $\lambda \phi^n$ theories in $d+1$D Minkowski spacetime, matched against the known renormalizability of the theories.}
\label{tab:summarytable}
\end{table}
%
\begin{align}
    I_{\vmodeone:\vmodetwo}
&\sim 
    \bigg[
        \hspace{-1pt}
        \prod_{j=3}^{n-1}
        \int_{\vec{z}_{j} \inn \mathbb{R}^{d}}  
        \hspace{-3pt}
    \bigg]
    \frac{
        r^{d-2} 
        z_{3}^{-1}
        \cdots
        z_{n-1}^{-1}
        \big|
            \vec{z}_{1} 
            + 
            \vec{z}_{2} 
            + 
            \vec{Z}
        \big|^{-1}
    }{
        \big[
            1
            + 
            r
            + 
            ||\vec{Z}||_{1}
            +
            \big|
                \vec{z}_{1} 
                + 
                \vec{z}_{2} 
                + 
                \vec{Z}
            \big|
        \big]^2
    }
    \,.
\end{align}
Noting that $ \vec{z}_{3}, \dots, \vec{z}_{n-1} $ are integrated over all momenta, we can rescale them as $ \vec{z}_{i} \to r \vec{z}_{i} $ without introducing an $ r $-dependence in the integrals. Further using $ \vec{z}_{2} = r \hat{z}_{2} $, we can perform an asymptotic expansion as $ r \to \infty $, to obtain
\begin{align}
\label{eq:twomodeMI}
    I_{\vmodeone:\vmodetwo}
&\sim 
    r^{-2[\lambda]}
    \,,
\end{align}
where $ [\lambda] = 2-\frac{1}{2}(d-1)(n-2) $ is the mass dimension of $ \lambda $. Note that this scaling is independent of the angle between $ \vmodeone $ and $ \vmodetwo $. The logarithmic derivative is now simply $\vphantom{\Big(} \Lim{r \to \infty} \frac{\d \ln \mathcal{I}_{r} (\infty)}{\d \ln r} = -2[\lambda]$, which reinforces the conclusions summarized in \cref{tab:summarytable}: $ \vphantom{\Big(} \Lim{r \to \infty} \frac{\d \ln \mathcal{I}_{r} (\infty)}{\d \ln r} < 0 $ for super-renormalizable theories, $ \vphantom{\Big(} \Lim{r \to \infty} \frac{\d \ln \mathcal{I}_{r} (\infty)}{\d \ln r} = 0 $ for marginal theories, and $ \vphantom{\Big(} \Lim{r \to \infty} \frac{\d \ln \mathcal{I}_{r} (\infty)}{\d \ln r} > 0 $ for non-renormalizable theories.

\section{Mutual information in de Sitter spacetime}
\label{sec:MIdS}

We next consider a massless Klein-Gordon field with a $ \lambda \phi^{n} $ interaction on a dynamical background, specifically the Poincar\'{e} patch of $d+1$D de Sitter spacetime, described by the action
\begin{align}
    S[\phi]
    =
    -
    \int \d^{d+1} x 
    \sqrt{-g} 
    \left[
        \frac{1}{2} (\del \phi)^{2} 
        + \frac{1}{2} \xi \mathcal{R} \phi^2 + \frac{\lambda}{n!} \phi^n
    \right]
    \,,
\end{align}
where $ (\del \phi)^2 = g^{\mu \nu} \del_{\mu} \phi \del_{\nu} \phi $ and $g = |g_{\mu \nu}|$ is the determinant of the metric. The de Sitter metric is defined by the line element $\d s^2 \equiv g_{\mu \nu} \d x^{\mu} \d x^{\nu} = -a^2 (\eta)[\d \eta^2 - \d \vx^2]$, written in terms of conformal time $\eta \in (-\infty, 0)$ and the scale factor $a(\eta) = -(H \eta)^{-1} $, where $H$ is the Hubble parameter. The Ricci scalar $\mathcal{R}$ is then given by $\mathcal{R} = 6 a''/a^3 = 12 H^2$, where primes indicate derivatives with respect to conformal time. We work with the standard canonically normalized field $\chi (\eta, \vx) = a^{(d-1)/2}(\eta) \phi(\eta, \vx) $, in which case, including the $ \sqrt{-g} $ from the measure, the time-dependence of the interaction strength is given by $ [a(\eta)]^{d+1} \lambda \phi^n  = \lambda_{d,n}(\eta) \chi^n  $, where
\begin{align}
    \lambda_{d,n}(\eta)
    =
    [a (\eta)]^{2 - \frac{1}{2}(d-1)(n-2)} 
    \lambda 
    \,.
\label{eq:dSlambda}
\end{align}
Since the mutual information is constructed from correlations of the interaction Hamiltonian, we are free to work entirely in the $ \chi $ basis.

We now quantize the field as usual and write the canonically normalized field operator in terms of its spatial Fourier modes as
\begin{align}
    \hat{\chi}(\eta, \vx)
    =
    \int_{\vk \inn \mathbb{R}^{d}} \hat{\chi}_{\vk}(\eta) e^{i \vk \cdot \vx}\,,
\end{align}
where the comoving momenta are defined as $\vk = a(\eta) \vk_{\text{ph}} $, and $ \hat{\chi}_{\vk}(\eta) = f_{k}^{>} (\eta) \hat{c}_{\vk} + f_{k}^{<} (\eta) \hat{c}_{-\vk}^{\dagger} $, with the mode functions now being solutions to the differential equation
\begin{align}
    \big(f_{k}^{\lessgtr}\big)''
    +
    \left[k^2 + \left(12 \xi - \frac{d^2 - 1}{4}\right)\frac{1}{\eta^2}\right]
    f_{k}^{\lessgtr}
= 
    0\,.
\end{align}
We now initialize the field in the Bunch-Davies vacuum at $ \eta_{0} \to - \infty $, and choose to conformally couple it to the de Sitter background by setting $ \xi = (d^2 - 1)/48 $, for which the mode functions are similar to those of Minkowski, that is 
\begin{align}
    f_{k}^{>}(\eta)
    =
    \frac{1}{\sqrt{2 \omega_k}} e^{- i \omega_k \eta}
    =
    f_{k}^{<*}(\eta)
    \,.
\label{eq:dSModeFunc}
\end{align}
Using these mode functions, we construct the propagators $ G^{>}_{k}(\eta_{1},\eta_{2}) =  f_{k}^{>}(\eta_{1}) f_{k}^{<}(\eta_{2}) $ and then, exchanging the time integrals over $ t_{1},t_{2} \in [t_0,t] $ for conformal time integrals over $ \eta_{1},\eta_{2} \in (-\infty,\eta] $, use \cref{eq:MIr} to compute the time-dependent mutual information density for $\lambda \phi^3$ and $\lambda \phi^4$ theories in $3+1$D de Sitter spacetime. From \cref{eq:dSlambda}, we see that in $3+1$D, the interaction strength diverges as $ \eta \to 0^- $ for $ n < 4 $, and thus perturbation theory is no longer valid. Therefore, we choose to normalize the mutual information for the $ \lambda \phi^{3} $ theory by $ \mathcal{I}_{1}(-H^{-1}) $, since $ \lambda_{d,n}(-H^{-1}) = \lambda $. On the other hand, for the $ \lambda \phi^{4} $ theory, the interaction strength in \cref{eq:dSlambda} is time-independent, and thus we can again normalize the mutual information by its late-time value, $ \mathcal{I}_{1}(0) $. 

\begin{figure*}
    \centering 
    \includegraphics[width=.495\textwidth]{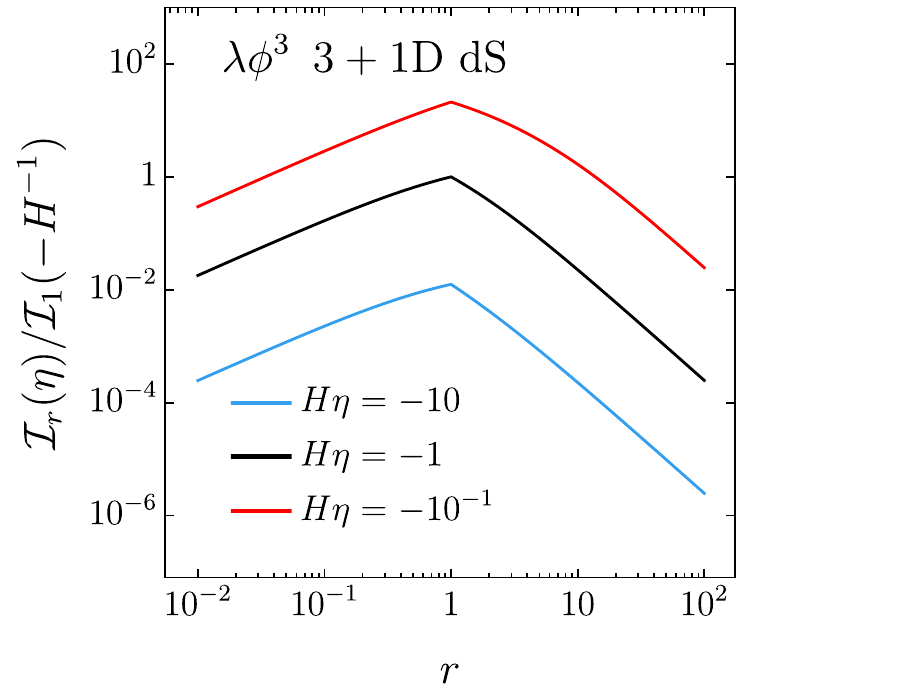}
    \includegraphics[width=.495\textwidth]{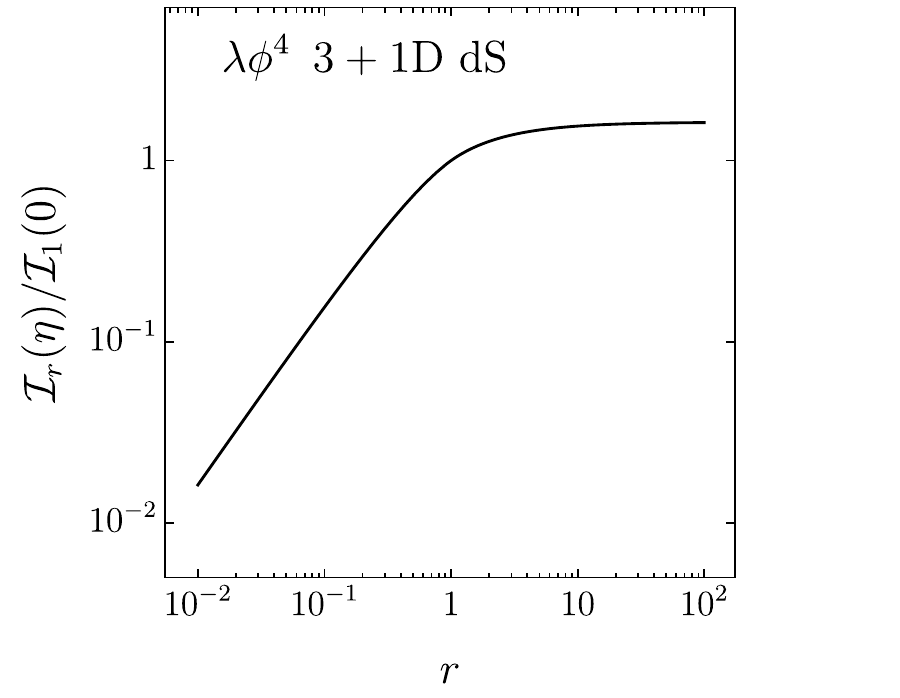}
    \caption{Time-dependent mutual information density as a function of mode separation in de Sitter spacetime for a (left) massless $\lambda \phi^3$ theory for $ \modeone = H $, normalized by the $r = 1$ result at $H \eta = -1$ and (right) massless $\lambda \phi^4$ theory for arbitrary $ \modeone $, normalized by the $r = 1$ result at $\eta \to 0^-$.}
    \label{fig:deSitter}
\end{figure*}

We show the resulting time-dependent mutual information for the massless $\lambda \phi^{3}$ theory in $3+1$D de Sitter spacetime in the left panel of \cref{fig:deSitter}. At each conformal time, we find that the mutual information decays as $ r \to \infty $, consistent with the late-time behavior in $3+1$D Minkowski spacetime shown in \cref{fig:TDTIminkowskiphi3}. As conformal time increases, we find that the mutual information increases nearly uniformly for all $ r $ and diverges as $ H \eta \to 0^- $, when perturbation theory is no longer valid. This breakdown for $ H \eta > -1 $ is signaled by a qualitative change in the behavior for $ r > 1 $, driven by terms that diverge as $ \ln(- H \eta) $. Nevertheless, we find that the logarithmic derivative remains well-defined and time-independent, with $\vphantom{\Big(} \Lim{r \to \infty} \frac{\d \ln \mathcal{I}_{r} (\eta)}{\d \ln r} = -2 $. Lastly, we show the results for the massless $\lambda \phi^{4}$ theory in $3+1$D de Sitter spacetime in the right panel of \cref{fig:deSitter}. In this case, we find that the mutual information density is time-independent and saturates to a finite value as $ r \to \infty $, with $\vphantom{\Big(} \Lim{r \to \infty} \frac{\d \ln \mathcal{I}_{r} (\eta)}{\d \ln r} = 0 $, consistent with the late-time behavior in $3+1$D Minkowski spacetime shown in \cref{fig:TDTIminkowskiphi4}.

\section{Discussion}
\label{sec:disc}

In this paper, we presented a mutual information-based measure to probe the renormalizability of QFTs both in and out of equilibrium. Specifically, we calculated the mutual information between two infinitesimal shells in momentum space, and showed that the logarithmic derivative, $\vphantom{\Big(} \Lim{r \to \infty} \frac{\d \ln \mathcal{I}_{r}}{\d \ln r}$, of the mutual information density $\mathcal{I}_{r}$ with mode separation $r$ is a reliable indicator of renormalizability.

We first considered $\lambda\phi^3$, $\lambda\phi^4$, and $\lambda\phi^6$ theories in different spatial dimensions in \\ Minkowski spacetime. We showed that, following an interaction quench, the late-time mutual information relaxes to that for the interacting vacuum and the logarithmic derivative of the late-time mutual information is negative for super-renormalizable theories, zero for renormalizable (marginal) theories, and positive for non-renormalizable theories. We further derived a scaling relation for the time-independent mutual information between two modes for $r \gg 1$ and showed that the logarithmic derivative is simply proportional to the mass dimension of $\lambda$ in this case, reaching the same general conclusions as the full shell calculation. We next considered conformally-coupled $\lambda\phi^3$ and $\lambda\phi^4$ theories on the Poincar\'{e} patch of $3+1$D de Sitter spacetime, initializing the field in the Bunch-Davies vacuum in the asymptotic past. We showed that at any finite time, the resulting mutual information has the same qualitative behavior as the late-time mutual information in Minkowski spacetime. We highlight that even when the time-dependence of the de Sitter background renders perturbation theory invalid, signaled by secular growth of the mutual information density at late times, the large-$r$ logarithmic slope continues to be well-defined and constant.

Our results elucidate the direct role that UV-IR correlations play in determining the renormalizability of interacting QFTs. They also suggest that the behavior of mutual information with mode separation is likely connected to RG flow and quantum criticality~\cite{Swingle:2010jz,Balasubramanian:2011wt,Beny:2012qh,Balasubramanian:2014bfa,Agon:2014uxa,Laflorencie:2015eck,Alves:2017fjk,Maity:2020qcg,FowlerIII:2020xgq,Fowler:2021oje,Paul:2022wgq,Cotler:2022fze,Zou:2022oxp,Costa:2022bvs,MartinsCosta:2022bjr,Kong:2025ksx}. In future work, it would be interesting to make this connection more precise and use mutual information to gain further insights into, for example,  the beta function, fixed points, and anomalous dimensions, perhaps by going to higher orders in perturbation theory -- see, for example, refs.~\cite{Ivanov:2024lbs,Kharuk:2025sqw} for higher order calculations. It would also be interesting to understand whether UV-IR correlations are directly related to mixed initial state counterterms as suggested in ref.\ \cite{Chaykov:2022pwd}, and how the measure defined here changes with the choice of initial state. Lastly, in the recent years, the open quantum system approach has led to various insights into interacting QFTs \cite{Lombardo:1995fg,Koksma:2009wa,Koksma:2011dy,Agon:2014uxa,Boyanovsky:2015xoa,Agon:2017oia,Boyanovsky:2018fxl,Burrage:2018pyg,Banerjee:2021lqu,Kading:2022jjl,Kading:2025cwg}, primordial perturbation theory \cite{Lombardo:2005iz,Boyanovsky:2015tba,Burgess:2015ajz,Hollowood:2017bil,Shandera:2017qkg,Boyanovsky:2018soy,Akhtar:2019qdn,Gong:2019yyz,Brahma:2020zpk,Banerjee:2020ljo,Brahma:2021mng,Hsiang:2021kgh,Colas:2022hlq,DaddiHammou:2022itk,Burgess:2022nwu,Colas:2022kfu,Burgess:2024eng,Salcedo:2024smn,Colas:2024ysu,Brahma:2024yor,Brahma:2024ycc,Lopez:2025arw,Li:2025azq,Salcedo:2025ezu,Burgess:2025dwm,Colas:2025app}, and holography \cite{Jana:2020vyx,Loganayagam:2020eue,Loganayagam:2022zmq,Pelliconi:2023ojb}, which, in turn, have led to important questions about renormalization in open QFTs \cite{Agon:2014uxa,Baidya:2017eho,Agon:2017oia,Burrage:2018pyg,Avinash:2019qga,Jana:2020vyx,Bowen:2024emo,Burgess:2024heo}. It would be interesting to use mutual information to probe renormalization in open QFTs and to extend this analysis to other theories where the renormalization picture may not be clear but the mutual information is still computable, to establish whether its behavior can serve as a fully action-independent diagnostic of renormalizability.

\acknowledgments{It is a pleasure to thank Yi-Zen Chu, Sachin Jain, Archana Kamal, and Sarah Shandera for useful discussions. This work was supported by the Department of Energy under award number DE-SC0020360. A.~F. was also supported by a NASA-Massachusetts Space Grant Consortium fellowship under award number 80NSSC20M0048 and by UML-Kennedy College of Sciences summer fellowships.}

\appendix

\section{General normal ordering}
\label{app:NormalOrdering}

In this appendix, we describe a general normal ordering prescription for bosonic operators such that correlators of the form $ \langle \normor{\hphi^n(x)} \normor{\hphi^m(y)} \rangle $ are free of contributions from bubbles (vacuum, thermal, etc.). When the system is initialized at zero temperature, the usual prescription -- that of Wick ordering -- is to arrange strings of creation and annihilation operators so that all annihilation operators are moved to the right-hand side. Although this procedure is efficient for the vacuum, it does not generalize straightforwardly to other commonly used states, such as Gaussian states. Therefore, it is advantageous to define normal ordered operators as a series of terms constructed to cancel such bubbles explicitly, 
\begin{align}
     \langle \normor{\hphi^n(x)} \normor{\hphi^m(y)} \rangle 
= 
    \langle \hphi^n(x) \hphi^m(y) \rangle - \text{(terms with bubbles)} 
    \,,
\end{align}
since it makes explicit what Wick ordering removes at the level of vacuum correlators and is readily extensible to other states. Note that while we use this definition for the vacuum in the main text, the construction presented here applies to any Gaussian state $ \hat{\rho}_{\text{G}} $. 

Bubbles arise in Wick's contractions when two field operators at the same spacetime event are contracted. To illustrate, consider a term in which $ n - 2 $ of the field operators at $ x $ are each contracted with fields at $y$, forming loops, while the last two $ \hphi(x) $ are contracted together, constituting a bubble,
\begin{align}
    \langle
        \hphi^n(x)
        \hphi^m(y)
    \rangle
\supset   
    \overbrace{
        \langle
            \hphi(x)
            \hphi(y)
        \rangle^{n-2}
    }^{\text{loops}}
    \langle
        \hphi^{m-n+2}(y)
    \rangle
    \overbrace{
        \langle
            \hphi^{2}(x)
        \rangle
    }^{\text{bubble}}
\,.
\label{eq:bubbleEXappA}
\end{align}
In fact, more bubbles will be generated in this term from $ \langle \hphi^{m-n+2}(y) \rangle $, but the important observation is that any number of field operators could be correctly ``loop-contracted'', but unless all are, there will be at least one bubble. Hence, we \textit{define} normal ordered products of field operators to be such that they can only be loop-contracted in correlators, and denote them with the usual notation, for example, $ \normor{\hphi^{n}} $. As a result, we can construct an operator that would reproduce \cref{eq:bubbleEXappA} when used in place of $ \hphi^{n}(x) $. In particular, $ \normor{\hphi^{n-2}(x)} \langle \hphi^{2}(x) \rangle $ results in
\begin{align}
    \langle
        [
            \normor{\hphi^{n-2}(x)} 
            \langle \hphi^{2}(x) \rangle
        ]
        \hphi^m(y)
    \rangle
&=
    \langle
        \normor{\hphi^{n-2}(x)} 
        \hphi^m(y)
    \rangle
    \langle \hphi^{2}(x) \rangle
\\
\nn
&=
    (n-2)!
    \langle
        \hphi(x)
        \hphi(y)
    \rangle^{n-2}
    \langle
        \hphi^{m-n+2}(y)
    \rangle
    \langle
        \hphi^{2}(x)
    \rangle
\,.
\end{align}
In this way, we can decompose products of field operators as linear combinations of the basis $ \{\langle \hphi^{2} \rangle^{n}\} $, as 
\begin{align}
    \hphi^{n}
&=
    c_{0}
    \normor{
        \hphi^{n}
    }
    + \,
    c_{1}
    \normor{
        \hphi^{n-2}
    }
    \langle
        \hphi^{2}
    \rangle
    +
    \cdots
    +
\begin{cases}
    c_{n/2}
    \langle
        \hphi^{2}
    \rangle^{n}
    &\,,
    \hspace{5pt}
    n
    \text{ even}
    \,,
\\
    c_{\lfloor n/2 \rfloor}
    \normor{
        \hphi
    }
    \langle
        \hphi^{2}
    \rangle^{\lfloor n/2 \rfloor}
    &\,,
    \hspace{5pt}
    n
    \text{ odd}
    \,.
\end{cases}
\end{align}
The combinatorial coefficients, $ \{c_{i}\} $, can be counted by considering the coefficient $ c_{k} $ on $ \normor{\hphi^{n-2 k}} \langle \hphi^{2} \rangle^{k} $. This term represents the case where only $ n-2k $ of the $ n $ fields would be loop-contracted. Hence, $ 2k $ of the field operators are contracted in $ k $ bubbles, that can happen in $ n $ choose $ 2k $ ways. We then multiply this by the number of ways to fully contract the $ 2k $ field operators into $k$ pairs, that can be counted by observing that
\begin{align}
    \underbrace{
        \hphi
        \hphi
        \cdots
        \hphi
    }_{2k}
    \to
    \langle
        \hphi^{2}
    \rangle
    \underbrace{
        \hphi
        \hphi
        \cdots
        \hphi
    }_{2k-2}
    \hspace{11.5pt}
    \text{ in }
    2k-1
    \text{ ways.}
\end{align}
Iterating this, we find that
\begin{align}
    \underbrace{
        \hphi
        \hphi
        \cdots
        \hphi
    }_{2k}
    \to
    \langle
        \hphi^{2}
    \rangle^{k}
    \hspace{11.5pt}
    \text{ in }
    (2k-1)!!
    \text{ ways.}
\end{align}
Therefore, $ c_{k} = \binom{n}{2k} (2k-1)!!  $ and $ \phi^{n} $ normal ordered with respect to $ \hat{\rho}_{\text{G}} $ is
\begin{align}
\label{eq:GeneralNormalOrder}
    \normor{
        \hphi^{n}
    }
&=
    \hphi^{n}
    -
    \sum_{k=1}^{\lfloor n/2 \rfloor}
    \binom{n}{2k}
    (2k-1)!!
    \normor{
        \hphi^{n-2k}
    }
    \langle
        \hphi^{2}
    \rangle^{k}
    \,.
\end{align}
We can quickly show that the expectation value of $ \normor{\hphi^{n}} $ vanishes in the state $ \hat{\rho}_{\text{G}} $ inductively. First, note that
\begin{align}
    \langle
        \normor{
            \hphi^{2}
        }
    \rangle
&=
    \langle
        \hphi^{2}
    \rangle
    -
    \langle
        \hphi^{2}
    \rangle
=
    0
    \,,
\end{align}
and 
\begin{align}
    \langle
        \normor{
            \hphi^{3}
        }
    \rangle
&=
    \langle
        \hphi^{3}
    \rangle
    -
    \binom{3}{2}
    \langle
        \normor{
            \hphi
        }
    \rangle
    \langle
        \hphi^{2}
    \rangle
=
    3
    \langle
        \hphi
    \rangle
    \langle
        \hphi^{2}
    \rangle
    -
    3
    \langle
        \hphi
    \rangle
    \langle
        \hphi^{2}
    \rangle
=  
    0
    \,.
\end{align}
Now assuming that this holds up to $ n - 1 $, we immediately find using \cref{eq:GeneralNormalOrder} that
\begin{align}
    \langle
        \normor{
            \hphi^{n}
        }
    \rangle
&=
\begin{cases}
    \langle
        \hphi^{n}
    \rangle
    -
    (n-1)!!
    \langle
        \hphi^{2}
    \rangle^{n/2}
=
    0
    &\,,
    \hspace{5pt}
    n
    \text{ even}
    \,,
\\
    \langle
        \hphi^{n}
    \rangle
    -
    n!!
    \langle
        \hphi
    \rangle
    \langle
        \hphi^{2}
    \rangle^{(n-1)/2}
=  
    0
    &\,,
    \hspace{5pt}
    n
    \text{ odd}
    \,,
\end{cases}
\end{align}
which can be verified using the same iterative counting as before. 

It is also straightforward to check that \cref{eq:GeneralNormalOrder} reproduces the usual normal ordering for the vacuum by expanding the $ \hphi^{n} $ in \cref{eq:GeneralNormalOrder} in terms of creation and annihilation operators and then using the commutation relations to move all of the annihilation operators to the right in each of the terms.

\section{General entanglement entropy calculation}
\label{app:entropycalc}

In this appendix, we derive the second order expansion of the entanglement entropy for the time-evolved free-theory vacuum under a quench Hamiltonian of the form in \cref{eq:IntHam}, introduced in \cref{sec:MomentumSpaceEE}. We follow a procedure similar to ref.~\cite{Balasubramanian:2011wt}: first, we determine the eigenvalues of the reduced density operator $\hat{\rho}_{X} $ of the system, and then we compute the entanglement entropy using the definition $S_{X} = -\text{Tr}[\hat{\rho}_{X} \ln \hat{\rho}_{X}]$. 

We begin by writing the state of the joint system as the ground state of the free theory time-evolved by the unitary operator of the full theory in the interaction picture
\begin{align}
    \ket{\Omega(t)}
    =
    \hat{U}(t,t_0)
    \ket{0_{X },0_{\bar{X} }}  \,.
\end{align}
Inserting a complete set of full system Fock states,
\begin{align}
    \Id 
&=
    \bkouter{\Sys{0},\Env{0}}{\Sys{0},\Env{0}}
    +
    \sum_{m\neq 0}
        \bkouter{m,\Env{0}}{m,\Env{0}}
    +
    \sum_{M\neq 0}
        \bkouter{\Sys{0},M}{\Sys{0},M}
    +
    \sum_{m,M \neq 0}
        \bkouter{m,M}{m,M}
    \,,
\end{align}
yields,
\begin{align}
    \ket{\Omega(t)}
&=
    Z(t) \ket{\Sys{0},\Env{0}}
    +
    \sum_{m \neq 0} A_m (t) \ket{m,\Env{0}}
    +
    \sum_{M \neq 0} B_M (t) \ket{\Sys{0},M}
    +
    \sum_{m,M \neq 0} C_{m,M} (t) \ket{m,M}
    \,,
\end{align}
where the coefficients are given by the amplitudes
\begin{align}
    Z(t) &= \braket{\Sys{0},\Env{0}|\hat{U}(t,t_0)|\Sys{0},\Env{0}} \,,
    \\
    A_{m}(t) &= \braket{m,\Env{0}|\hat{U}(t,t_0)|\Sys{0},\Env{0}} \,,
    \\
    B_{M}(t) &= \braket{\Sys{0},M|\hat{U}(t,t_0)|\Sys{0},\Env{0}} \,,
    \\
    C_{m,M}(t) &= \braket{m,M|\hat{U}(t,t_0)|\Sys{0},\Env{0}}
    \,.
\end{align}
We obtain the reduced density matrix of the system from the joint system's density matrix $\hat{\rho}_{X \bar{X} }(t) = \bkouter{\Omega(t)}{\Omega(t)}$ by tracing over the environment,
\begin{align}
    \Sys{\hat{\rho}}(t) 
    &= 
    \Env{\text{Tr}}[\hat{\rho}_{X \bar{X} }(t)]
    =
	\begin{bmatrix}
		\begin{array}{c|c}
			\mathcal{C}(t)
			&
			\begin{matrix}
				\phantom{.}
				&
				\mathcal{V}^{\dagger}(t)
				&
				\phantom{.}
			\end{matrix}
			\\
			\hline
			\begin{matrix}
				\phantom{.}
				\\
				\mathcal{V}(t)
				\\
				\phantom{.}
			\end{matrix}
			&
			\begin{matrix}
				\phantom{.}
				&
				\phantom{.}
				&
				\phantom{.}
				\\
				\phantom{.}
				&
				\mathcal{M}(t)
				&
				\phantom{.}
				\\
				\phantom{.}
				&
				\phantom{.}
				&
				\phantom{.}
			\end{matrix}
		\end{array}
	\end{bmatrix}
	\,,
\end{align}
where
\begin{align}
\label{eq:curlyC}
	\mathcal{C}(t)
&=
	|Z(t)|^2
	+
	\sum_{M \neq 0}
	|B_{M}(t)|^2
=
    1
    -
    \text{Tr}
	\mathcal{M}(t)
	\,,
\\
\label{eq:curlyV}
	\mathcal{V}_{m}(t)
&=
	A_m(t)
	Z^*(t)
	+
	\sum_{M \neq 0}
	C_{m, M}(t)
	B_M^*(t)
	\,,
\\
\label{eq:curlyM}
	\mathcal{M}_{mn}(t)
&=
	A_m(t)
	A_n^*(t)
	+
	\sum_{M \neq 0}
	C_{m, M}(t)
	C_{n, M}^*(t)
	\,,
\end{align}
The second equality in \cref{eq:curlyC} follows from the conservation of probability, that is, \\ $ \text{Tr}_{X}[\hat{\rho}_{X}(t)] = 1 $. 

We denote the $i^{\text{th}}$ eigenvalue of $\hrho_X(t)$ with $\rho_i(t)$, and write the eigenvalue equation 
\begin{align}
	\left[
		\hat{\rho}_X(t)
		-
		\rho_i(t)
		\Id
	\right]
	\ket{\psi_{i}}
=
	\ket{\text{null}}
	\,,
\end{align}
In block matrix form, the operator on the left-hand side is
\begin{align}
	\left[
		\hat{\rho}_X(t)
		-
		\rho_i(t)
		\Id
	\right]
=
	\begin{bmatrix}
		\begin{array}{c|c}
			\mathcal{C}(t)
			-
			\rho_i(t)
			&
			\begin{matrix}
				\phantom{.}
				&
				\mathcal{V}^{\dagger}(t)
				&
				\phantom{.}
			\end{matrix}
			\\
			\hline
			\begin{matrix}
				\phantom{.}
				\\
				\mathcal{V}(t)
				\\
				\phantom{.}
			\end{matrix}
			&
			\begin{matrix}
				\phantom{.}
				&
				\phantom{.}
				&
				\phantom{.}
				\\
				\phantom{.}
				&
				\mathcal{M}(t)
				-
				\rho_i(t)
                \Id_{\setminus 0}
				&
				\phantom{.}
				\\
				\phantom{.}
				&
				\phantom{.}
				&
				\phantom{.}
			\end{matrix}
		\end{array}
	\end{bmatrix}
	\,,
\label{eq:rhoXeigvaleq}
\end{align}
where $ \Id_{\setminus 0} $ is the identity with the dimensions of $ \mathcal{M}(t) $. We would now like to compute $ \det ( \hat{\rho}_X(t) - \rho_i(t) \Id) = 0 $ perturbatively. To do so, we use Schur's formula\footnote{(Schur's formula) Given a block matrix $ M $ of the form
\begin{align}
    M
    =
    \begin{bmatrix}
        A & B \\ C & D
    \end{bmatrix}
\end{align}
where $ A $ is $ m \times m $ and \textit{invertible}, $ B $ is $ m \times n $, $ C $ is $ n \times m $, and $ D $ is $ n \times n $, 
\begin{align}
    \det(M)
    =
    \det(A)
    \det(D - C A^{-1} B)
    \,.
\end{align}}
to take the determinant of \cref{eq:rhoXeigvaleq}, obtaining the characteristic equation
\begin{align}
	\left[
        \mathcal{C}(t)
		-
        \rho_i(t)
	\right]
	\det
	\left[
		\mathcal{M}(t)
		-
        \rho_i(t)
        \Id_{\setminus 0}
        -
        \mathcal{V}(t)
        \frac{
            1
        }{
            \mathcal{C}(t)
            -
            \rho_i(t)
        }
        \mathcal{V}^{\dagger}(t)
	\right]
=
	0
	\,.
\end{align}
Then, we use the second order expansion of the time-evolution operator
\begin{align}
    \hat{U}(t,t_0)
=
    \Id 
    -
    i 
    \int_{t_1}
    \hat{H}_{\text{I}}(t_1)
    -
    \frac{1}{2}
    \int_{t_1}
    \int_{t_2}
    \hspace{0.75pt}T\hspace{-0.75pt}
    \big\{
    \hat{H}_{\text{I}}(t_1)
    \hat{H}_{\text{I}}(t_2)
    \big\}
    \,,
\end{align}
to write 
\begin{align}
    Z(t) 
&= 
    \braket{\Sys{0},\Env{0}|\hat{U}(t,t_0)|\Sys{0},\Env{0}}
=
	1
    +
    Z^{(2)}(t) 
    +
    O(\lambda^3)
    \,,
\\
    A_{m}(t) 
&= 
    \braket{m,\Env{0}|\hat{U}(t,t_0)|\Sys{0},\Env{0}}
=
    A_{m}^{(2)}(t) 
    +
    O(\lambda^3)
    \,,
\\
    B_{M}(t) 
&= 
    \braket{\Sys{0},M|\hat{U}(t,t_0)|\Sys{0},\Env{0}}
=
    B_{M}^{(2)}(t) 
    +
    O(\lambda^3)
    \,,
\\
\label{eq:CmMExpansion}
    C_{m,M}(t) 
&= 
    \braket{m,M|\hat{U}(t,t_0)|\Sys{0},\Env{0}}
=
    C_{m,M}^{(1)}(t) 
    +
    C_{m,M}^{(2)}(t) 
    +
    O(\lambda^3)
    \,,
\end{align}
where we used that the interaction Hamiltonian is normal ordered with respect to the vacuum of the free theory to drop the first order terms in $ Z(t) $, $ A_m(t) $, and $ B_M(t) $. Extending this to the density matrix blocks, we have
\begin{align}
	\mathcal{C}(t)
&=
    1
    -
    \sum_{m \neq 0}
    \sum_{M \neq 0}
    |C_{m, M}^{(1)}(t)|^{2}
    +
    O(\lambda^3)
	\,,
\\
	\mathcal{V}_{m}(t)
&=
    A_{m}^{(2)}(t) 
    +
    O(\lambda^3)
	\,,
\\
	\mathcal{M}_{mn}(t)
&=
	\sum_{M \neq 0}
	C_{m, M}^{(1)}(t)
	C_{n, M}^{(1)*}(t)
    +
    O(\lambda^3)
	\,.
\end{align}
Now, restricting the eigenvalues $\rho_i$ to second order, the characteristic equation becomes
\begin{align}
	\left[
        1
        -
        \text{Tr}
        \mathcal{M}^{(2)}(t)
		-
        \rho_i(t)
	\right]
	\det
	\left[
        \mathcal{M}^{(2)}(t)
		-
        \rho_i(t)
        \Id_{\setminus 0}
	\right]
=
	0
	\,,
\end{align}
and we conclude that the first eigenvalue of $ \hat{\rho}_X(t) $ must be $1 - \text{Tr} \mathcal{M}^{(2)}(t)$, while the remaining eigenvalues must match those of the matrix $ \mathcal{M}^{(2)}(t) $. Therefore, denoting the $i^{\text{th}}$ eigenvalue of $ \mathcal{M}^{(2)}(t) $ by $ \bar{\lambda}^2 a_i $ -- where we have absorbed the dimensions of $ \lambda $ in $ a_{i} $, so that both $ \bar{\lambda} $ and $ a_{i} $ are dimensionless -- we define a matrix with the same eigenvalues as $ \hat{\rho}_{X} $,
\begin{align}
    \hat{\rho}_{X,\lambda}
=
    \bigg[
        1
        -
        \bar{\lambda}^2
        \sum_{i \neq 0}
        a_i
    \bigg]
    \ket{\psi_0}
    \bra{\psi_0}
    +
    \bar{\lambda}^2
    \sum_{i \neq 0}
    a_i
    \ket{\psi_i}
    \bra{\psi_i}
    \,,
\end{align}
where the $ \{\ket{\psi_i}\} $ form an orthonormal basis. 

Now, we can compute the entanglement entropy for system $X$ as
\begin{align}
	S_X
&=
	-
	\bigg(
		1
		-
            \bar{\lambda}^2
		\sum_{i \neq 0}
		a_{i}
	\bigg)
	\ln
	\bigg(
		1
		-
		\bar{\lambda}^{2}
		\sum_{j \neq 0}
		a_{j}
	\bigg)
	-
	\sum_{i \neq 0}
	\bar{\lambda}^{2}
	a_{i}
	\ln
	\left(
		\bar{\lambda}^{2}
		a_{i}
	\right)
\nn \\
&=
	-
	\bigg(
		1
		-
		\bar{\lambda}^{2}
		\sum_{i \neq 0}
		a_{i}
	\bigg)
	\ln
	\bigg(
		1
		-
		\bar{\lambda}^{2}
		\sum_{j \neq 0}
		a_{j}
	\bigg)
	-
	\bar{\lambda}^{2}
	\ln
	\left(
		\bar{\lambda}^{2}
	\right)
	\sum_{i \neq 0}
	a_{i}
	-
    \bar{\lambda}^{2}
	\sum_{i \neq 0}
	a_{i}
	\ln
	\left(
		a_{i}
	\right)
\nn \\
&=
	-
	\bar{\lambda}^{2}
	\ln
	\left(
		\bar{\lambda}^{2}
	\right)
	\sum_{i \neq 0}
	a_{i}
	+
	\bar{\lambda}^{2}
	\sum_{i \neq 0}
	a_{i}
	\left(
		1
		-
		\ln
		a_{i}
	\right)
    +
    O(\bar{\lambda}^{3})
	.
\end{align}
Thus, to leading order, the entanglement entropy is 
\begin{align}
\label{eq:EEwithC2}
    \Sys{S}(t) 
    =
    -
    \ln (\bar{\lambda}^2) 
    \text{Tr}
    \mathcal{M}^{(2)}(t)
    =
    -
    \ln (\bar{\lambda}^2) 
    \sum_{m,M\neq0} 
    |C_{m,M}^{(1)}(t)|^2
    \,.
\end{align}
Using the definition of $C^{(1)}_{m,M}(t)$ from \cref{eq:CmMExpansion} and that $ \sum_{m \neq 0} \bkouter{m}{m} = \Id - \bkouter{0}{0} $, we find that for an interaction-picture interaction Hamiltonian of the form
\begin{align}
    \hat{H}_{\text{I}}(t)
    =
    \lambda
    \sum_{\alpha=0}^{n}
        \int \d^d x
        \SyO_{n-\alpha} (t, \vx)
        \EnO_\alpha (t, \vx)
    \,,
\end{align}
the sum over $ |C^{(1)}_{m,M} (t)|^2 $ in \cref{eq:EEwithC2} can be written as
\begin{align}
    \sum_{m,M\neq0} 
    |C_{m,M}^{(1)}(t)|^2
&=
    \lambda^2
    \sum_{\alpha, \beta}
        \int_{x_1, x_2}
            \hspace{-3pt}
            \braket
                {
                \Sys{0}|
                \SyO_{n-\alpha}(x_1)
                \SyO_{n-\beta}(x_2)
                |\Sys{0}
                }_c
            \braket
                {
                \Env{0}|
                \EnO_{\alpha}(x_1)
                \EnO_{\beta}(x_2)
                |\Env{0}
                }_c
    \,,
\label{eq:C2asOcorrs}
\end{align}
where $x_i \equiv (t_i, \vx_{i})$, $\int_{x_i} \equiv \int_{t_0}^{t} \d t_i \int \d^d x_i$, and we have defined the connected correlators $\braket{0|\SyO_1 \SyO_2|0}_c = \braket{0| \SyO_{1} \SyO_{2} |0} - \braket{0| \SyO_{1} |0} \braket{0| \SyO_{2} |0}$, which arise because of the identity insertion for both the system and environment Fock bases. Note that, because the interaction Hamiltonian is normal ordered, all one-point functions vanish, so the subscript $c$ in the correlator is retained only for completeness. We now use \cref{eq:C2asOcorrs} in \cref{eq:EEwithC2} to write the entanglement entropy in the final form
\begin{align}
    S_{X}(t)
    =
    -
    \lambda^2 \ln \bar{\lambda}^2
    \sum_{\alpha, \beta}
        \int_{x_1,x_2}
            \braket
                {
                \Sys{0}|
                \SyO_{n-\alpha}(x_1)
                \SyO_{n-\beta}(x_2)
                |\Sys{0}
                }_c
            \braket
                {
                \Env{0}|
                \EnO_{\alpha}(x_1)
                \EnO_{\beta}(x_2)
                |\Env{0}
                }_c
    \,,
\label{eq:EEITOCCsAppB}
\end{align}
which matches \cref{eq:SvN}.

\section{\texorpdfstring{Entanglement entropy density for $\lambda \phi^n$ in $d+1$D}{Entanglement entropy density for λϕn in d + 1D}}
\label{app:EEndDerivation}

In this appendix, we derive the entanglement entropy density $ S_{X} / L^{d} $ for a $\lambda \phi^n$ theory in $d+1$D Minkowski spacetime. We recall that the bipartite entanglement entropy of region $ X $ is given by \cref{eq:SvN},
\begin{align}
    S_{X}(t)
&=
    -\lambda^2 \ln \bar{\lambda}^2
    \sum_{\alpha, \beta}
        \int_{x_1,x_2}
            \braket
                {
                \Sys{0}|
                \SyO_{n-\alpha} (x_1)
                \SyO_{n-\beta} (x_2)|
                \Sys{0}
                }_c
            \braket
                {
                \Env{0}|
                \EnO_\alpha (x_1)
                \EnO_\beta (x_2)|
                \Env{0}
                }_c
    ,
\label{eq:EEITOCCsAppC}
\end{align}
where, for the normal ordered $ \lambda \normor{\hat{\phi}^{n}}$ interaction, partitioning the field as $ \hphi = \hphi_{X} + \hphi_{\bar{X}} $ results in the operators 
\begin{align}
    \SyO_{n-\alpha} (x)
    &\equiv 
    \frac{
        1
    }{
        (n-\alpha)!
    }
    \normor{ \hat{\phi}_{X}^{n-\alpha} (x) }
    \,,
\label{eq:OAppC}
\\
    \EnO_\alpha(x) 
    &\equiv 
    \frac{
        1
    }{
        \alpha!
    }
    \normor{\hat{\phi}_{\bar{X}}^\alpha (x)}
    \,.
\label{eq:barOAppC}
\end{align}
From the prescription in \cref{app:NormalOrdering}, the normal ordering in \cref{eq:OAppC,eq:barOAppC} requires that each of the field operators in $ \SyO_{n-\alpha} (x_1) $ is contracted with those in $ \SyO_{n-\beta} (x_2) $, and similarly for $ \EnO_\alpha (x_1) $ and $\EnO_\beta (x_2) $. Thus, all terms with $ \alpha \neq \beta $ in \cref{eq:EEITOCCsAppC} vanish. Further, since
\begin{align}
    \braket{\Sys{0}|
        \SyO_{0} (x_1)
        \SyO_{0} (x_2)
    |\Sys{0}}_c
    &\equiv 
    \braket{\Sys{0}|
        \normor{ \hat{\phi}_{X}^{0} (x_1) } 
        \normor{ \hat{\phi}_{X}^{0} (x_2) } 
    |\Sys{0}}_c
    =
    \braket{\Sys{0}|
        \Id_{X}
    |\Sys{0}}_c
    =
    0
    \,,
\\
    \braket{\Env{0}|
        \EnO_{0} (x_1)
        \EnO_{0} (x_2)
    |\Sys{0}}_c
    &\equiv 
    \braket{\Sys{0}|
        \normor{ \hat{\phi}_{\bar{X}}^{0} (x_1) } 
        \normor{ \hat{\phi}_{\bar{X}}^{0} (x_2) } 
    |\Sys{0}}_c
    =
    \braket{\Env{0}|
        \Id_{\bar{X}}
    |\Env{0}}_c
    =
    0
    \,,
\end{align}
we can also drop the $\alpha = 0$ and $\alpha = n$ terms. Then using \cref{eq:OAppC,eq:barOAppC}, we find that
\begin{align}
    S_{X}(t)
&=
    -
    \lambda^2 \ln \bar{\lambda}^2
    \sum_{\alpha = 1}^{n-1}
    \frac{
        1
    }{
        \left[
            \alpha!
            \left(
                n
                -
                \alpha
            \right)!
        \right]^{2}
    }
\nn \\
&\qquad \times
    \int_{x_1, x_2}
    \braket
        {
        \Sys{0}|
        \normor{ \hat{\phi}_{X}^{n-\alpha} (x_1)}
        \normor{ \hat{\phi}_{X}^{n-\alpha} (x_2)}
        |\Sys{0}
        }_c
    \braket
        {
        \Env{0}|
        \normor{\hat{\phi}_{\bar{X}}^\alpha (x_1)}
        \normor{\hat{\phi}_{\bar{X}}^\alpha (x_2)}
        |\Env{0}
        }_c
        \,.
\end{align}
We can next use Wick's contractions to write
\begin{align}
    \braket{0|
        \normor{ \hat{\phi}^n (x_1) }
        \normor{ \hat{\phi}^n (x_2) }
    |0}_c
&=
    n!
    \prod_{i=1}^{n}
    \int_{\vk_{i} \inn \mathbb{R}^{d}} 
    G_{k_i}^{>} (t_1, t_2) 
    e^{i \vk_{i} \cdot (\vx_{1} - \vx_{2})}
    \,,
\end{align}
where we defined the propagator $G^{>}_{k_i} (t_1, t_2) \equiv f_{k_i}^{>}(t_1) f_{k_i}^{<} (t_2)$. The entanglement entropy then becomes
\begin{align}
\label{eq:EntangleEntX}
    S_{X}(t)
    &=
    -
    \sum_{\alpha = 1}^{n-1}
    \frac{
    \lambda^2 \ln \bar{\lambda}^2
    }{
        \alpha!
        \left(
            n
            -
            \alpha
        \right)!
    }
    \int_{t_1, t_2}
    \Bigg[
        \prod_{i=1}^{n-\alpha}
        \prod_{j=n-\alpha+1}^{n}
        \int_{\vk_{i} \inn X, \vk_{j} \inn \bar{X}} 
        G_{k_i}^{>}(t_1, t_2) 
        G_{k_j}^{>}(t_1, t_2) 
    \Bigg]
\nn \\
&\qquad \times
    \int_{\vx_1, \vx_2}
    e^{
        i 
        \sum_{m=1}^{n}
        \vk_{m} 
        \cdot 
        (\vx_{1} - \vx_{2})
    }
    \,.
\end{align}
Diagrammatically, the contributions to the entanglement entropy can be represented as
\begin{align}
\begin{tikzpicture}
    \node at (-75 pt, 0) {\LARGE $ \sum\limits_{\alpha=1}^{n-1} \int_{t_1,t_2} $};
    \node at (-38 pt,0) {$t_{1}$};
     \draw (-30 pt,0) parabola[parabola height= 4 pt] (30 pt,0);
     \draw (-30 pt,0) parabola[parabola height= 8 pt] (30 pt,0);
     \draw (-30 pt,0) parabola[parabola height= 12 pt] (30 pt,0);
     \node at (0, 15 pt) {$\cdot$};
     \node at (0, 17 pt) {$\cdot$};
     \node at (0, 19 pt) {$\cdot$};
     \draw (-30 pt,0) parabola[parabola height= 22 pt] (30 pt,0);
     \draw (-30 pt,0) parabola[parabola height= 26 pt] (30 pt,0);
     \draw[thick,decorate,decoration={brace}] (50 pt, 26 pt) -- (50 pt,0);
     \node[right] at (55 pt, 13 pt) {$n-\alpha$ system propagators};
     \node at (38 pt,0) {$t_{2}$};
     \draw[dashed] (-30 pt,0) parabola[parabola height= -4 pt] (30 pt,0);
     \draw[dashed] (-30 pt,0) parabola[parabola height= -8 pt] (30 pt,0);
     \draw[dashed] (-30 pt,0) parabola[parabola height= -12 pt] (30 pt,0);
     \node at (0, -15 pt) {$\cdot$};
     \node at (0, -17 pt) {$\cdot$};
     \node at (0, -19 pt) {$\cdot$};
     \draw[dashed] (-30 pt,0) parabola[parabola height= -22 pt] (30 pt,0);
     \draw[dashed] (-30 pt,0) parabola[parabola height= -26 pt] (30 pt,0);
     \draw[thick,decorate,decoration={brace}] (50 pt, 0) -- (50 pt,-26 pt);
     \node[right] at (55 pt, -13 pt) {$\alpha$ environment propagators};
\end{tikzpicture}
\end{align}
where solid and dashed lines represent system and environment propagators, respectively. Let us next consider the spatial integrals of \cref{eq:EntangleEntX}, that can be evaluated as 
\begin{align}
\label{eq:EntangleVol}
    \int_{\vx_1, \vx_2}
    e^{
        i 
        \sum_{m=1}^{n}
        \vk_{m} 
        \cdot 
        (\vx_{1} - \vx_{2})
    }
=
    (2 \pi)^{d}
    \delta^{d}
    \left(
        \mathlarger{\textstyle \sum_{m=1}^{n}} \vk_{m}
    \right)
    \int
    \d^d x_{1}
    \,.
\end{align}
Since the remaining integral produces a volume factor,\footnote{A more rigorous way to obtain the volume factor is to first place the system in a finite-size box and then take the infinite-size limit at the end of the calculation.} we divide \cref{eq:EntangleEntX} by $ L^{d} $, and work with the entanglement entropy \textit{density}. Now adding and subtracting terms corresponding to $ \alpha = 0 $ and $ \alpha = n $ in \cref{eq:EntangleEntX}, the summation becomes a binomial expansion, that can be written as 
\begin{align}
\label{eq:EntangleBinomial}
    \sum_{\alpha = 0}^{n}
    \binom{n}{\alpha}
    \left[
        \int_{\vk \inn X} 
        G_{k}^{>}(t_1, t_2) 
    \right]^{n-\alpha}
    \left[
        \int_{\vk \inn \bar{X}} 
        G_{k}^{>}(t_1, t_2) 
    \right]^{\alpha}
=
    \bigg[
        \int_{\vk \inn \mathbb{R}^{d}} 
        G_{k}^{>}(t_1, t_2) 
    \bigg]^{n}
    \,,
\end{align}
where we used $ X \cup \bar{X} = \mathbb{R}^{d} $. Using \cref{eq:EntangleVol,eq:EntangleBinomial} in \cref{eq:EntangleEntX}, and subtracting out the $\alpha = 0$ and $\alpha = n$ terms, the entanglement entropy density finally becomes
\begin{align}
    \frac{S_{X} (t)}{L^d}
=&
    -\frac{\lambda^2 \ln \bar{\lambda}^2}{n!} 
    \int_{t_1, t_2}
    \bigg[
        \prod_{i=1}^{n}
            \int_{\vk_{i} \in \mathbb{R}^d} 
        -
        \prod_{i=1}^{n}
            \int_{\vk_{i} \in X} 
        -
        \prod_{i=1}^{n}
            \int_{\vk_{i} \in \bar{X}} 
    \bigg]
\nn \\
&\quad \times
    (2 \pi)^{d}
    \delta^{d}
    \left(
        \mathlarger{\textstyle \sum_{m=1}^{n}} \vk_{m}
    \right)
    \prod_{j=1}^{n}
        G_{k_j}^{>} (t_1, t_2)
    \,,
\label{eq:infvolEEapp}
\end{align}
which matches \cref{eq:infvolEE}.

\section{\texorpdfstring{Mutual information density for $\lambda \phi^n$ in $d+1$D}{Mutual information density for
λϕn in d + 1D}}
\label{app:ndderivation}

In this appendix, we derive the leading-order contribution to the mutual information density between two infinitesimally thin shells $A$ and $B$ around the momentum scales $ k_{A} $ and $ k_{B} $ shown in \cref{fig:subspaces2D}, that is
\begin{align}
    A 
&= 
    \big\{\vk : k_{A} - \frac{\delta k}{2} \leq k \leq  k_{A} + \frac{\delta k}{2} \big\}
    \,,
\nn \\
    B 
&= 
    \big\{\vk : k_{B} - \frac{\delta k}{2} \leq k \leq  k_{B} + \frac{\delta k}{2} \big\} 
    \,,
\end{align}
where $\delta k \ll \min (k_{A}, k_{B}) $. 

We begin by recalling the general expression for the mutual information between regions $ A $ and $ B $ written in terms of entanglement entropies,
\begin{align}
\label{eq:MIappC}
    I_{A:B} = S_{A} + S_{B} - S_{A \cup B}
    \,.
\end{align}
We now partition the full momentum space into three regions: two thin shells $ A $ and $ B $, and the environment $ E = \mathbb{R}^{d} \setminus (A \cup B) $. For each of the entropies $ S_{A} $, $ S_{B} $, and $ S_{A \cup B} $, the regions $X$ and  $ \bar{X} $ are thus different; in particular, 
\begin{align}
    S_A &: X  = A,\, \bar{X}  = B \cup E
    \,,
    \\
    S_B &: X  = B,\, \bar{X}  = A \cup E
    \,,
   \\ 
    S_{A \cup B} &: X  = A \cup B,\, \bar{X}  = E
    \,.
\end{align}
Since we found that the entanglement entropy must be defined as a density, we similarly work with the mutual information density, $ I_{A:B}/L^{d} $. Now using \cref{eq:infvolEEapp} in \cref{eq:MIappC}, we find that the mutual information density can be written in terms of integrals over $A$, $B$, and $\mathbb{R}^d$ as
\begin{align}
    \frac{I_{A:B} (t)}{L^d}
&=
    -\frac{\lambda^2 \ln \bar{\lambda}^2}{n!} 
    \int_{t_1, t_2}
    \int_{A:B}
    \prod_{j=1}^{n}
        G_{k_j}^{>} (t_1, t_2)
    \,,
\end{align}
where we defined the shorthand 
\begin{align}
    \int_{A:B}
\equiv &
    \left[
        \prod_{i=1}^{n}
            \int_{\vk_{i} \inn \mathbb{R}^d} 
        -
        \prod_{i=1}^{n}
            \int_{\vk_{i} \inn A} 
        -
        \prod_{i=1}^{n}
            \int_{\vk_{i} \inn B} 
        +
        \prod_{i=1}^{n}
            \int_{\vk_{i} \inn A \cup B} 
        -
        \prod_{i=1}^{n}
            \int_{\vk_{i} \inn \bar{A}} 
        -
        \prod_{i=1}^{n}
            \int_{\vk_{i} \inn \bar{B}} 
        +
        \prod_{i=1}^{n}
            \int_{\vk_{i} \inn E} 
    \right]
\nn \\
&\quad \times
    (2 \pi)^{d}
    \delta^{d}
    \left(
        \mathlarger{\textstyle \sum_{m=1}^{n}} \vk_{m}
    \right)
    \,.
\label{eq:intABappC}
\end{align}
Since regions $ A $ and $ B $ are shells of infinitesimal thickness, we can next assume that the mutual information density is approximately uniform throughout the shells and evaluate integrals over $ A $ (or $B$, with $A \to B$) as
\begin{align}
    \int_{\vk_{i} \inn A} 
    f(\vk_{i},\cdots)
=
    \int_{\Omega_{i}}
    \int_{\modeone -\delta k/2}^{\modeone + \delta k/2}
    \frac{
        \d k_{i}
    }{
        (2 \pi)^{d}
    }
    k_{i}^{d-1}
    f(k_{i}, \Omega_{i}, \cdots)
=
    \frac{
        \delta k
        \modeone^{d-1}
    }{
        (2 \pi)^{d}
    }
    \int_{\Omega_{i}}
    f(\modeone, \Omega_{i}, \cdots)
    \,,
\end{align}
where $ \int_{\Omega_{i}} $ is an integral over the solid angle of dimension $d-1$, corresponding to $ \vk_{i} $. With each integral over $A$ or $B$ producing a factor of $ \delta k $, we can expand the mutual information density as a series in $ \delta k $. Further assuming $ A \cap B = \emptyset $, so that the following identities hold,
\begin{align}
    \int_{\vk \inn A \cup B} 
= 
    \int_{\vk \inn A} + \int_{\vk \inn B}
\,,
\hspace{12pt}
    \int_{\vk \inn \bar{A}} 
= 
    \int_{\vk \inn B} + \int_{\vk \inn E}
\,,
\hspace{6pt}
\text{ and }
\hspace{6pt}
    \int_{\vk \inn \bar{B}} 
= 
    \int_{\vk \inn A} + \int_{\vk \inn E}
\,,
\label{eq:InteIdentApp}
\end{align}
we can rewrite \cref{eq:intABappC} as
\begin{align}
    \int_{A:B}
&\approx
    n(n-1)
    \int_{\vk_{1} \inn A} 
    \int_{\vk_{2} \inn B} 
    \left[
        \prod_{i=3}^{n}
            \int_{\vk_{i} \inn \mathbb{R}^d}
    \right]
    (2 \pi)^{d}
    \delta^{d}
    \left(
        \mathlarger{\textstyle \sum_{m=1}^{n}} \vk_{m}
    \right)
    +
    O(\delta k^{3})
    \,.
\end{align}
Keeping just the leading-order contribution, the mutual information density between momentum scales $ \modeone $ and $ \modetwo $ can now be written as
\begin{align}
    \frac{I_{\modeone:\modetwo} (t)}{L^d}
&=
    -\frac{\lambda^2 \ln \bar{\lambda}^2}{(n-2)!} 
    \int_{t_1, t_2}
    \int_{\vk_{1} \inn A} 
    \int_{\vk_{2} \inn B} 
    \left[
        \prod_{i=3}^{n}
            \int_{\vk_{i} \inn \mathbb{R}^d} 
    \right]
    (2 \pi)^{d}
    \delta^{d}
    \left(
        \mathlarger{\textstyle \sum_{m=1}^{n}} \vk_{m}
    \right)
    \prod_{j=1}^{n}
        G_{k_j}^{>} (t_1, t_2)
    \,.
\end{align}
Finally, defining the ratio $ r = \modetwo / \modeone $, denoting the mutual information density \\ $  I_{\modeone:r\modeone} (t) / L^d \equiv \mathcal{I}_{r}(t) $, and evaluating the radial parts of the $ \vk_{1} $ and $ \vk_{2} $ integrals, we obtain
\begin{align}
    \mathcal{I}_{r}(t)
=&
    -
    \frac{
        \lambda^2 \ln \bar{\lambda}^2
        \delta k^{2}
        \modeone^{2d-2}
        r^{d-1}
    }{
        (2 \pi)^{2d}
        (n-2)!
    }
\int_{t_1, t_2}
    G_{\modeone}^{>}(t_1, t_2)
    G_{r\modeone}^{>}(t_1, t_2)
\nn \\
&\quad
    \times
    \int_{\Omega_{1}, \Omega_{2}}
    \left[
        \prod_{i=3}^{n}
            \int_{\vk_{i} \inn \mathbb{R}^d} 
            G_{k_{i}}^{>} (t_1, t_2)
    \right]
    (2 \pi)^{d}
    \delta^{d}
    \big(
        \vec{K}_{r}^{(n)}
    \big)
    \,,
\end{align}
where $ \vec{K}_{r}^{(n)} \equiv \modeone(\hat{k}_{1} + r \hat{k}_{2}) + \sum_{m=3}^{n} \vk_{m} $, with $ \hat{k}_{1} $ and $ \hat{k}_{2} $ being unit vectors, which matches \cref{eq:MIr}.

\section{\texorpdfstring{Integrals in $d+1$D $ \lambda \phi^{n} $ }{Integrals in d + 1-dimensional λϕn}}
\label{app:calcspecs}

In this appendix, we provide some mathematical details for computing the integrals in the mutual information density of \cref{eq:MIr}. For simplicity, we write \cref{eq:MIr} as 
\begin{align}
\label{eq:MIrApp}
    \mathcal{I}_{r}(t)
&=
    -
    \frac{
        \lambda^2 \ln \bar{\lambda}^2
        \delta k^{2}
        \modeone^{2d-2}
        r^{d-1}
    }{
        (2 \pi)^{2d}
        (n-2)!
    }
    \int_{t_1, t_2}
    G_{\modeone}^{>}(t_1, t_2)
    G_{r\modeone}^{>}(t_1, t_2)
    \mathcal{J}_{r}(t_1, t_2)
    \,,
\end{align}
and work with the quantity
\begin{align}
\label{eq:Jfunction}
    \mathcal{J}_{r}(t_1, t_2)
=
    \int
    \d^{d-1} \Omega_{1}
    \int
    \d^{d-1} \Omega_{2}
    \left[
        \prod_{i=3}^{n-1}
            \int 
            \frac{\d^{d} k_{i}}{(2 \pi)^{d}}
            G_{k_{i}}^{>} (t_1, t_2)
    \right]
    G_{|\vec{K}_{r}^{(n-1)}|}^{>} (t_1, t_2)
    \,,
\end{align}
where $ \vec{K}_{r}^{(n)} \equiv \modeone(\hat{k}_{1} + r \hat{k}_{2}) + \sum_{m=3}^{n} \vk_{m} $, the solid angle is
\begin{align}
    \d^{d-1} \Omega
&= 
    \d\phi
    \left[
        \prod_{p=1}^{d-2}
        \d\theta_{p} 
        \sin^{p} \theta_{p}
    \right]
    \,,
\end{align}
where $\phi \in [0,2\pi)$ and $\theta_p \in [0,\pi)$, and the differential volume element in $d$ dimensions is
\begin{align}
    \d^d k
    = 
    k^{d-1}
    \d k \,
    \d^{d-1} \Omega
    \,.
\end{align}
In the remainder of this appendix, we describe how to compute \cref{eq:Jfunction} for specific $n$ and $d$.

\subsection{\texorpdfstring{$\lambda \phi^3$ in $d+1$D}{λϕ3 in d + 1D}}
\label{app:phi3cal}

For $ n = 3 $, we can compute all integrals for general $d$, and all but one of the angular integrals are trivial. Using 
\begin{align}
    \int
    \d^{d-1} \Omega
=
    \frac{2 \pi^{d/2}}{\Gamma(d/2)}
    \,,
\end{align}
\cref{eq:Jfunction} reduces to
\begin{align}
\label{eq:Jfunction3}
    \mathcal{J}_{r}(t_1, t_2)
=
    \frac{(2 \pi)^{d}}{\pi \Gamma(d-1)}
    \int_{0}^{\pi}
    \d\theta
    \sin^{d-2} \theta
    \, G_{\modeone\sqrt{1 + r^2 + 2 r \cos \theta}}^{>} (t_1, t_2)
    \,,
\end{align}
where $ \theta $ is the angle between $\hat{k}_{1}$ and $\hat{k}_{2}$. We now make the standard substitution $u = \cos\theta$, to find 
\begin{align}
    \mathcal{J}_{r}(t_1, t_2)
=
    \frac{(2 \pi)^{d}}{\pi \Gamma(d-1)}
    \int_{-1}^{1} 
    \d u 
    [1-u^2]^{(d-3)/2} 
    G_{\modeone\sqrt{1 + r^2 + 2 r u}}^{>} (t_1, t_2)
    \,.
\end{align}
This remaining integral is now computable -- for instance using \texttt{Mathematica} -- and we use it in \cref{eq:MIrApp} to obtain the results shown in \cref{fig:TDTIminkowskiphi3}.

\subsection{\texorpdfstring{$\lambda \phi^4$ in $d+1$D}{λϕ4 in d + 1D}}
\label{app:phi4cal}

\subsubsection*{\texorpdfstring{$\lambda \phi^4$ in $2+1$D}{2+1D}}

For $ n = 4 $ and $ d = 2 $, \cref{eq:Jfunction} becomes
\begin{align}
\label{eq:Jfunctionr42}
    \mathcal{J}_{r}(t_1, t_2)
=&
    \, \frac{1}{2 \pi}
    \int_{0}^{2 \pi}
    \d \phi_{2}
    \int_{0}^{2 \pi}
    \d \phi_{3}
    \int_{0}^{\infty} 
    \d k_{3}
    \, k_{3} 
    G_{k_{3}}^{>} (t_1, t_2)
    G_{|\modeone(\hat{k}_{1} + r \hat{k}_{2}) + \vk_{3}|}^{>} (t_1, t_2)
    \,,
\end{align}
where $ \phi_{2} $ is the angle between $ \hat{k}_{1} $ and $ \hat{k}_{2} $, $ \phi_{3} $ is the angle between $ \hat{k}_{1} $ and $ \vk_{3} $, and  $|\modeone(\hat{k}_{1} + r \hat{k}_{2}) + \vk_{3}| = 
        [\modeone^{2}
        (1 + 2 r \cos\phi_{2} 
        +
        r^{2})
        +
        2 \modeone k_{3}
        \{\cos \phi_{3} + r \cos(\phi_{2} - \phi_{3})\}
        +
        k_{3}^{2}]^{1/2}
    $. We perform the radial integral $\int_{0}^{\infty} \d k_3$ analytically, and numerically integrate the remaining shell integrals over $\phi_2$ and $\phi_3$. We use this strategy to compute \cref{eq:MIrApp} and produce the $2+1$D result shown in \cref{fig:TDTIminkowskiphi4}.

\subsubsection*{\texorpdfstring{$\lambda \phi^4$ in $3+1$D}{3+1D}}

For $ n = 4 $ and $ d = 3 $, \cref{eq:Jfunction} becomes
\begin{align}
\label{eq:Jfunctionr43}
    \mathcal{J}_{r}(t_1, t_2)
=
    \frac{1}{\pi}
    \int_{0}^{\pi}
    \d\theta_{2} 
    \sin \theta_{2}
    \int 
    \d^{3} k_{3}
    G_{k_{3}}^{>} (t_1, t_2)
    G_{|\modeone(\hat{k}_{1} + r \hat{k}_{2}) + \vk_{3}|}^{>} (t_1, t_2)
    \,,
\end{align}
where $ \theta_{2} $ is the angle between $ \hat{k}_{1} $ and $ \hat{k}_{2} $. In this case, we perform the change of variables
\begin{align}
    \vq_{2} 
&\equiv 
    \modeone
    \big(
        \hat{k}_{1} 
        +
        r 
        \hat{k}_{2}
    \big)
    \,,
\nn \\
    \vq_{3} 
&\equiv 
    \vq_{2} 
    + 
    \vk_{3}
    \,,
\end{align}
whose magnitudes can be written as
\begin{align}
    q_2 
&= 
    \modeone
    \left(
        1 + r^2 + 2 r \cos\theta_2 
    \right)^{1/2} 
    \in 
    \left[
        \modeone|1 - r|, \modeone(1 + r)
    \right]
    \,,
\\
    q_3 
&= 
    \left(
        q_2^2 + k_3^2 + 2 q_2 k_3 \cos\theta_3 
    \right)^{1/2} 
    \in 
    \left[
        |q_2 - k_3|, q_2 + k_3
    \right]
    \,,
\end{align}
where $\theta_3$ is the angle between $\vq_{2}$ and $\vk_{3}$. Using these, we can write
\begin{align}
    \cos\theta_2 &= \frac{q_2^2}{2 r \modeone^{2}} - \frac{1}{2 r} - \frac{r}{2}\,,
\\
    \cos\theta_3 &= \frac{q_3^2}{2 q_2 k_3} - \frac{q_2}{2 k_3} - \frac{k_3}{2 q_2} \,,
\end{align}
and hence,
\begin{align}
    -\sin\theta_2 \d \theta_2 &= \frac{q_2}{r \modeone^{2}} \d q_2\,,
\\
    -\sin\theta_3 \d \theta_3 &= \frac{q_3}{q_2 k_3} \d q_3 \,.
\end{align}
Therefore, \cref{eq:Jfunctionr43} becomes
\begin{align}
    \mathcal{J}_{r}(t_1, t_2)
&=
    \frac{2}{r \modeone^{2}}
    \int_{0}^{\infty}
    \d k_3
    \, k_3 
    \int_{\modeone|1 - r|}^{\modeone(1 + r)}
    \d q_2
    \int_{|q_2 - k_3|}^{q_2 + k_3}
    \d q_3
    \, q_3 \, 
    G_{k_3}^{>} (t_1, t_2)
    G_{q_3}^{>} (t_1, t_2)
    \,.
\end{align}
These integrals can now be computed directly and we use the result in \cref{eq:MIrApp} to produce the $3+1$D result shown in \cref{fig:TDTIminkowskiphi4}.

\subsubsection*{\texorpdfstring{$\lambda \phi^4$ in $5+1$D}{5+1D}}

For $ n = 4 $ and $ d = 5 $, \cref{eq:Jfunction} becomes
\begin{align}
\label{eq:Jfunctionr45}
    \mathcal{J}_{r}(t_1, t_2)
=
    \frac{1}{6 \pi}
    \int_{0}^{\pi}
    \d\theta_{2} 
    \sin^{3} \theta_{2}
    \int 
    \d^{5} k_{3}
    G_{k_{3}}^{>} (t_1, t_2)
    G_{|\modeone(\hat{k}_{1} + r \hat{k}_{2}) + \vk_{3}|}^{>} (t_1, t_2)
    \,,
\end{align}
where $ \theta_{2} $ is the angle between $ \hat{k}_{1} $ and $ \hat{k}_{2} $. We perform the same change of variables as in the $ n = 4 $ and $ d = 3 $ case and regulate the $ k_{3} $ integral using dimensional regularization, letting $ k_{3}^{4} \to k_{3}^{4+\xi} $. We use the result in \cref{eq:MIrApp}, taking the late-time limit of the time integrals as described in the main text, to obtain the $5+1$D result shown in \cref{fig:TDTIminkowskiphi4}.

\bibliography{references}
\bibliographystyle{JHEP}

\end{document}